\documentclass[10pt,a4paper]{article}
\usepackage[latin1]{inputenc}
\usepackage{amsmath}
\usepackage{amsfonts}
\usepackage{amssymb}
\usepackage{graphics}
\usepackage{psfrag}
\usepackage{epsfig}
\usepackage{pstricks}

\newcommand{\nn}{\nonumber}
\newcommand{\ph}[1]{\phantom{#1}}

\newcommand{\db}{D^{(\bar{A})}}

\author{
Hans  Westman $^1$, Tom  Z\l o\'{s}nik $^2$\\
{\small \it $(1)$ Instituto de F\'{i}sica Fundamental, CSIC,
Serrano 113-B, 28006 Madrid, Spain},\\
{\small \it $(2)$ Imperial College Theoretical Physics, Blackett Laboratory, London, SW7 2AZ, United Kingdom}}

\date{\today}
\title{Exploring Cartan gravity with dynamical symmetry breaking}
\begin{document}
\maketitle

\begin{abstract}
It has been known for some time that General Relativity can be regarded as a Yang-Mills-type gauge theory in a symmetry broken phase. In this picture the gravity sector is described by an $SO(1,4)$ or $SO(2,3)$ gauge field $A^{a}_{\ph{a}b\mu}$ and Higgs field $V^{a}$ which acts to break the symmetry down to that of the Lorentz group $SO(1,3)$. This symmetry breaking mirrors that of electroweak theory. However, a notable difference is that while the Higgs field $\Phi$ of electroweak theory is taken as a genuine dynamical field satisfying a Klein-Gordon equation, the gauge independent norm $V^2\equiv \eta_{ab}V^{a}V^{b}$ of the Higgs-type field $V^a$ is typically regarded as non-dynamical. Instead, in many treatments $V^a$ does not appear explicitly in the formalism or is required to satisfy 
$V^2 = \mathrm{const.} \neq 0$ by means of a Lagrangian constraint. As an alternative to this we propose a class of polynomial actions that treat both the gauge connection $A^{a}_{\ph{a}b\mu}$ and Higgs field $V^a$ as genuine dynamical fields with no {\em ad hoc} constraints imposed. The resultant equations of motion consist of a set of first-order partial differential equations. We show that for certain actions these equations may be cast in a second-order form, corresponding to a scalar-tensor model of gravity. One simple choice leads to the extensively studied Peebles-Ratra rolling quintessence model. Another choice yields a scalar-tensor symmetry broken phase of the theory with positive cosmological constant and an effective mass $M$ of the gravitational Higgs field ensuring the constancy of $V^2$ at low energies and agreement with empirical data if $M$ is sufficiently large. More general cases are discussed corresponding to variants of Chern-Simons modified gravity and scalar-Euler form gravity, each of which yield propagating torsion.
\end{abstract}

\section{Introduction}
The question of whether the gravitational field is essentially similar to other fields in nature or not has been a long running theme in physics. In particular, how far do the commonalities between gravitation and the gauge theories of particle physics extend?  The basic ingredients of these gauge theories are gauge fields which are one-forms valued in the Lie algebra of some group $G$. In gauge theories with a spontaneously broken symmetry there are, in addition to gauge fields, Higgs fields which are space-time scalar fields belonging to some representation of $G$ which may break the symmetry $G$ down to a residual symmetry $H\subset G$ at the level of the equations of motion via the attainment of non-vanishing vacuum expectation values. An example of this, in the context of the standard model of particle physics, is the electroweak theory. In that model the gauge fields are a $U(1)$ gauge field $B\equiv B_{\mu}dx^{\mu}$ coupling to hypercharge and an $SU(2)$ gauge field $C^{{\cal A}}_{\ph{{\cal A}}{\cal B}} \equiv C^{{\cal A}}_{\ph{{\cal A}}{\cal B}\mu}dx^{\mu}$ coupling to isospin, where ${\cal A}$ and ${\cal B}$ are $SU(2)$ indices. These fields are accompanied by a $U(1)\times SU(2)$ vector field $\Phi^{\cal A}$, called the electroweak Higgs field. When $\Phi^{\dagger}\Phi \equiv \left<\Phi^{\dagger}\right>\left<\Phi\right>\neq 0$ the $SU(2)\times U(1)$ gauge symmetry is broken leaving a remnant $U(1)$ symmetry preserving the form of $\Phi^{\cal A}$. In this example then the group $G$ is identified with $U(1)\times SU(2)$ and $H$ is identified with the $U(1)$ symmetry of electromagnetism.

We now consider the gravitational field. This field is typically described entirely by a rank-2 space-time tensor $g_{\mu\nu}$ referred to as the metric tensor. We will refer to this as the second-order formulation of gravity. In the sense of the above definitions, this field is neither a gauge field nor a Higgs field \footnote{In the second-order formulation of gravity, the Christoffel connection is indeed a $GL(4)$ connection but it is constructed from $g_{\mu\nu}$ and therefore not usually thought of as the fundamental variable.}.  However, the metric formulation of gravity may be indeed be recovered from ingredients that have more, but not everything, in common with gauge theory. This is possible via the first-order Einstein-Cartan formulation of gravity \cite{Kibble:1961ba} wherein gravity is described by an $SO(1,3)$ gauge field $\omega^{I}_{\ph{I}J}\equiv \omega^{I}_{\ph{I}J\mu}dx^{\mu}$, referred to as the spin-connection, and a Lorentz-vector and one-form $e^{I}\equiv e^{I}_{\mu}dx^{\mu}$, referred to as the co-tetrad. From the Einstein-Cartan perspective, gravitation is described by gauge invariant actions with the Lorentz group $SO(1,3)$ as the local symmetry group $G$. The metric tensor is identified as $g_{\mu\nu}=\eta_{IJ}e^{I}_{\mu}e^{J}_{\nu}$ where $\eta_{IJ}=\mathrm{diag}(-1,1,1,1)$ is the invariant $SO(1,3)$ matrix. While the connection $\omega^{I}_{\ph{I}J}$ has all the mathematical characteristics of a standard gauge field the co-tetrad $e^I$ does not appear to have a counterpart in gauge theory \footnote{A similar remark applies to the Palatini first-order formulation in which the dynamical variables are the $GL(4)$  gauge field $\Gamma^\rho_{\mu\nu}$, the affine connection, and the metric $g_{\mu\nu}$, which again does not appear to have a natural counterpart in gauge theory.}. It is an  $SO(1,3)$ vector (and so possesses an `internal group' index)
but also possesses a space-time index $\mu$, a combination not found in any other known fundamental field in physics. This dissimilarity with gauge theory may ultimately be superficial if it turns out that the co-tetrad is a composite object formed of more familiar objects. For example, an approach which has attracted recent interest is to retain the $SO(1,3)$ gauge field and regard $e^{I}$ as just such a composite object. For instance \cite{Akama:1978pg,Diakonov:2011im,Obukhov:2012je} have considered models involving sets of $spin(1,3)$ spinor fields $\{\Psi_{(i)}\}$ where $e^{I}_{\mu}\equiv i\sum_{(i)}\left(\bar{\Psi}_{(i)}\gamma^{I}D_{\mu}\Psi_{(i)}-D_{\mu}\bar{\Psi}_{(i)}\gamma^{I}\Psi_{(i)}\right)$.

An alternative to the above spinorial approach is to regard the $SO(1,3)$ symmetry of the Einstein-Cartan theory as the remnant symmetry $H\subset G$ after spontaneous symmetry breaking, i.e. we begin with a symmetry group larger than $SO(1,3)$ and via symmetry breaking retain only local Lorentz symmetry. The ingredients necessary for this will thus be standard quantities from gauge theory: a gauge field $A^{a}_{\ph{a}b}\equiv A^{a}_{\ph{a}b\mu}dx^{\mu}$ and a Higgs field $V^{a}$ where $a,b,..$ are gauge indices of the larger group. The larger group can be taken to be the de Sitter $SO(1,4)$ group, the anti-de Sitter $SO(2,3)$, or the Poincare group $ISO(1,3)$. This approach is that of Cartan geometry \cite{Wise:2009fu} where the gauge field and Higgs field admit a simple geometrical interpretation in terms of `idealized waywisers' \cite{Westman:2012xk}. The gauge connection $A^{a}_{\ph{a}b }$  dictates how much a symmetric space (either the de-Sitter, anti-de Sitter, or Minkowski spacetime) is rotated when 
parallel transported (`rolling without slipping') along some path on the spacetime manifold. The Higgs field $V^a$ corresponds to the `point of contact' between the symmetric space and the manifold. For concreteness we restrict attention to cases where the larger group is either of the $10$-parameter groups $SO(1,4)$ and $SO(2,3)$.

If we are interested in eliminating differences between gravity and the gauge theories of particle physics, this approach is promising as one may use the above ingredients to construct an object which, like the co-tetrad, possesses a space-time index and a gauge index. This is simply the covariant derivative of $V^{a}$:
\begin{eqnarray}
D_{\mu}V^{a} \equiv  \partial_{\mu}V^{a}+A^{a}_{\ph{a}b \mu}V^{b}. \label{dvo}
\end{eqnarray}
By construction, this quantity transforms as a one-form under coordinate transformations and as a vector under $SO(1,4)$ or $SO(2,3)$ (which we will henceforth write more compactly as $SO(1,4)|SO(2,3)$) transformations. Consider a field $V^{a}$, where $a,b,c,..$ now specifically refer to $SO(1,4)|SO(2,3)$ indices. If $V^{2}\equiv \eta_{ab}V^{a}V^{b} = - \ell^{2}$ for $SO(2,3)$ and $V^{2}=\ell^{2}$ for $SO(1,4)$ then a gauge may be found where $V^{a}\doteq\ell \delta^{a}_{4}$ (the notation $\doteq$ denotes an equation holding in a particular preferred gauge), where $\ell$ is some real non-zero constant. One may check that the generators of $SO(1,4)|SO(2,3)$ which leave the above form of $V^{a}$ invariant are those of the Lorentz group $SO(1,3)$. Therefore, for the symmetry breaking ansatz $V^{2}= \mp \ell^{2}$ we have $G=SO(1,4)|SO(2,3)$ and $H=SO(1,3)$.  We will refer to any indices projected along $V^{a}$ as $4$ and the remaining $SO(1,3)$ indices as $I,J..$. Given the ansatz for $V^{2}$ we have from (\ref{dvo}) that: 
\begin{eqnarray}
D_{\mu}V^{a}  &\doteq& A^{a}_{\ph{a}4\mu}\ell. \label{dv}
\end{eqnarray}
It can be seen that if $V^{2}=\mathrm{const.}$ (where the constant is of appropriate sign) then $D_{\mu}V^{a}$ only contains non-vanishing components orthogonal to $V^{a}$ and transforms homogeneously under the $SO(1,3)$ transformations defined by $V^{a}$ i.e. for constant $V^{2}$, the form $D_{\mu}V^{a}$ behaves like a co-tetrad. For non-constant but non-vanishing $V^{2}$ (though still of appropriate sign always), one may introduce a `projector' 
\begin{equation}\label{projector}
{\mathcal P}^{a}_{\phantom{a}b}\equiv \delta^{a}_{\phantom{a}b}-V^{a}V_{b}/V^{2}
\end{equation}
which acts to select components of $SO(1,4)|SO(2,3)$ tensors orthogonal to $V^{a}$. Hence one may define an object $E^{a}\equiv {\mathcal P}^{a}_{\phantom{a}b} D_{\mu}V^{a}$ which will transform as a co-tetrad.  Regarding the remaining parts of the curvature $A^{a}_{\ph{a}b}$, it additionally follows that $A^{I}_{\ph{I}J}$ has the mathematical properties of the spin-connection $\omega^{I}_{\ph{I}J}$. In these senses then, the familiar objects of Einstein-Cartan theory may be 
recovered in the symmetry broken phase of a $SO(1,4)|SO(2,3)$ gauge theory. No objects beyond a gauge field and a Higgs field have been introduced, in particular not additional metric or affine structure.

We now highlight the similarities between the electroweak theory and Cartan gravity. The following table displays a side-by-side comparison of the basic dynamical fields of electroweak theory and Cartan gravity with $SO(1,4)$ gauge group and with $V^2>0$. The remnant (stabilizer) symmetry group is defined by the subgroup that leaves the Higgs field invariant. In the case of electroweak theory the remnant symmetry group is the $U(1)_{EM}$ group of electromagnetism and in Cartan gravity it is $SO(1,3)$ local Lorentz invariance of the tangent spaces.
\begin{table}[h!]\label{EWCGcomp}
\begin{center}
\begin{tabular}{|l|l|l|}
\hline
& {\bf Electroweak theory} & {\bf Cartan gravity}\\
\hline Gauge connection: & ${W_\mu}^{\mathcal A}_{\ph A\mathcal B}(x)$ & ${A_\mu}^a_{\ph ab}(x)$\\
\hline Higgs field: & $\Phi^{\mathcal A}(x)$ & $V^a(x)$\\
\hline Symmetry group: & $U(1)_Y\times SU(2)$& $SO(1,4)|SO(2,3)$\\
\hline Stabilizer group: & $U(1)_{EM}$& $SO(1,3)$\\
\hline
\end{tabular}
\end{center}
\end{table}

Though this construction seems promising, it must be shown that $SO(1,4)|SO(2,3)$ invariant actions built from the pair $\{A^{a}_{\ph{a}b},V^{a}\}$ exist that can reduce to Einstein-Cartan gravity, and thus General Relativity, after symmetry breaking.
This was first shown to be possible by Stelle and West \cite{Stelle:1979va,Stelle:1979aj} and later by Pagels \cite{Pagels:1983pq}. Due to the relationship between the variables $\{A^{a}_{\ph{a}b},V^{a}\}$ and Cartan geometry \cite{SharpeCartan,Wise:2006sm}, we refer to gravity seen through these variables as Cartan Gravity.

In the papers \cite{Stelle:1979va,Stelle:1979aj,Pagels:1983pq} the recovery of the co-tetrad via (\ref{dv}) is aided by actually enforcing the symmetry breaking ansatz $V^{2}=\mp \ell^{2}$ by adding a Lagrange multiplier to the gravitational action. 
By analogy, in the electroweak theory this would be akin to fixing $\Phi^{\dagger}\Phi =\mathrm{const.}$ via a similar constraint. It appears to be the case however that fluctuations of the scalar $\Phi^{\dagger}\Phi$, in the form of the electroweak Higgs particle, have been detected. Therefore in electroweak theory, variations in the action $\Phi^{{\cal A}}$ are not restricted. If indeed Cartan gravity represents a step closer to the remaining parts of physics, it would seem that fluctuations in $\eta_{ab}V^{a}V^{b}$ should similarly be allowed.\footnote{The Higgs field also plays a central role in the renormalizability of the electroweak theory. Whether the introduction of the gravitational Higgs field $V^a$ can play a similar role should be investigated but is outside the scope of this paper.}

This implies that symmetry breaking solutions to a Cartan gravity model should follow from free variation of both  $A^{ab}$ and $V^a$. It is useful to take a step back to realize why it is in fact rather challenging to do so. Typically when one thinks of dynamics of a scalar degree of freedom $\varphi$, one typically imagines being able to write down a kinetic term $g^{\mu\nu}\partial_{\mu}\varphi\partial_{\nu}\varphi\sqrt{-g}$. We have seen that a `co-tetrad' $E^{a}_{\mu}\equiv {\mathcal P}^{a}_{\ph{a}b}D_{\mu}V^{b}$ may be constructed in an $SO(1,4)|SO(2,3)$ covariant sense. Hence one may define a metric $\bar{g}_{\mu\nu} \equiv  \eta_{ab}E^{a}_{\mu}E^{b}_{\nu}$ and inverse metric. It would seem then that one could simply write down a Klein-Gordon type Lagrangian under the identification $g_{\mu\nu}=\bar{g}_{\mu\nu}$ and $\varphi=V^{2}$.  This approach was advocated in \cite{Stelle:1979aj}. Note, however, that the construction of an action using the variable $\bar{g}_{\mu\nu}$, itself containing the projector ${\mathcal P}^{
a}_{\ph{a}b}$, represents an assumption at the level of the action $V^{2}\neq 0$. It would seem then that such actions are of limited use when looking for a model that can equally accommodate phases where $SO(1,4)|SO(2,3)$ gauge invariance is or is not spontaneously broken.

In contrast, in this paper we shall restrict attention to action principle which are polynomial in both $V^a$, $A^{ab}$, and their derivatives thereby circumventing the above-mentioned problem. By allowing {\em free, unconstrained} variation of both connection $A^{ab}$ and symmetry breaking field $V^a$ we effectively introduce a new scalar gauge-independent degree of freedom $V^2(x)$. It is then necessary to ask what that new degree of freedom could represent in our universe. From the point of view of cosmology it is natural to suspect that this new degree of freedom could serve as a dark energy candidate. Indeed, as we shall see, the extensively studied rolling quintessence model due to Peebles and Ratra \cite{Ratra:1987rm,Peebles:1987ek} corresponds to a very simple action (containing only two polynomial terms) with equations of motion obtained by unconstrained variation with respect to both connection and symmetry breaking field. This demonstration is non-trivial since the equations of motion that follows from our polynomial class of actions are necessarily first order partial differential equations. It is only after reformulating the theory in second order `metric' language when it becomes clear that the model is surprisingly that of Peebles-Ratra rolling quintessence. We also demonstrate how a standard type symmetry breaking mechanism can be obtained by including additional polynomial terms.

The plan of this paper is as follows: In Section \ref{poly} we exhibit the most general polynomial Cartan gravity action. In Section \ref{decomp} we discuss a choice of variables that enables the general action of \ref{poly} to be written in a particularly transparent form and facilitates reformulating the first-order theory in second-order form. In \ref{spece} we consider a subset of this general action and show that these types of theories are equivalent to a class of scalar-tensor theories. As may be expected, the scalar degree of freedom is encoded in the norm $V^{2}$.  In Section \ref{general} we consider the physical interpretation of more general actions, and in Section \ref{vnull} we discuss behaviour in the event that $V^{2}=0$. In Section \ref{discussion} we discuss the results of the previous sections.

Throughout the paper we work in units in which $c$ and $\hbar$ are dimensionless and numerically equal to one. Duration is measured in metres $m$ and mass in inverse metres $m^{-1}$.
\section{Polynomial actions for gravity}
\label{poly}
In accordance with the discussion in the previous section we will look to construct gravitational actions from the pair $\{A^{a}_{\ph{a}b},V^{a}\}$ which are each allowed to vary freely and assumed to satisfy $\delta A^{a}_{\ph{a}b}=\delta V^{a}=0$ on the boundary. As a further guide, we will additionally look to construct only  actions that are invariant under $SO(1,4)|SO(2,3)$ and polynomial in the basic fields $\{A^{a}_{\ph{a}b},V^{a}\}$. We shall see that these restrictions still allow a non-trivial collection of possible terms in the action and we shall here consider the most general combination of them. 

Given our restrictions, what terms may be considered? In general these terms will be actions (defined as integrals over space-time four-forms) built from the invariants $\eta_{ab},\epsilon_{abcde}$, the field $V^{a}$, and the curvature two-form of $A^{a}_{\ph{a}b}$ defined 
as $F^{a}_{\ph{a}b}=\frac{1}{2}F^{a}_{\ph{a}b\mu\nu}dx^{\mu}\wedge dx^{\nu}$ where $F^{a}_{\ph{a}b\mu\nu}\equiv 2\partial_{[\mu|}A^{a}_{\ph{a}b|\nu]}+2A^{a}_{\ph{a}c[\mu|}A^{c}_{\ph{c}b|\nu]}$. The $SO(1,4)| SO(2,3)$ indices may be raised with $\eta^{ab}$ and lowered with $\eta_{ab}$. For notational compactness we denote the wedge product $y\wedge z$ between differential forms $y$ and $z$ simply as 
$yz$. For example, if $y$ is a one-form and $z$ is a three-form then we have:

\begin{eqnarray}
\int yz &=& \int y \wedge z=\frac{1}{3!} \int y_{\mu}z_{\nu\sigma\delta} dx^{\mu}\wedge dx^{\nu}\wedge dx^{\sigma}\wedge dx^{\delta}\nn \\
   &=&  \frac{1}{3!}\int \varepsilon^{\mu\nu\sigma\delta}y_{\mu}z_{\nu\sigma\delta}  d^{4}x
\end{eqnarray}
where $\varepsilon^{\mu\nu\sigma\delta}$ is the Levi-Civita density. For an exposition of the calculus of variations using differential forms we refer to \cite{Westman:2012xk}. The most general action consistent with our requirements is as follows:
\begin{eqnarray}
\nn S[V^{a},A^{ab}]&=&\int \left(\alpha_{1} \epsilon_{abcde}V^{e}+\alpha_{2}V_{a}V_{c}\eta_{bd}+\alpha_{3}\eta_{ac}\eta_{bd}\right)F^{ab}F^{cd} \\
&&+\left(\beta_{1} \epsilon_{abcde}V^{e}+\beta_{2}V_{a}V_{c}\eta_{bd}+\beta_{3}\eta_{ac}\eta_{bd}\right)DV^{a}DV^{b}F^{cd}\nn \\
&&+\gamma_{1} \epsilon_{abcde}V^{e}DV^{a}DV^{b}DV^{c}DV^{d} \label{act1}
\end{eqnarray}
The $\alpha$,$\beta$, and $\gamma$ may themselves be polynomial functions of $V^{2}$. 
The action (\ref{act1}) may look unfamiliar. It contains terms quadratic in the curvature $F^{ab}$ 
but terms contributing to the equations of motion necessarily couple to the gravitational Higgs field $V^{a}$; the only such term which may be independent of $V^{a}$ is the $\alpha_{3}$ term. However, if $\alpha_{3}$ is constant then the term is simply proportional to $\int F_{ab}F^{ab}= \int d (A^{ab}F_{ab}+\frac{1}{3}A^{ac}A_{a}^{\ph{a}d}A_{cd})$ and so may be neglected as a boundary term.

\section{Decomposing the connection} 
\label{decomp}
The equations of motion deduced from a variational principle of polynomial actions written in terms of forms and gauge covariant derivatives can be shown to invariably be first-order partial differential equations \cite{Westman:2012zk}. Thus, the above polynomial action (\ref{act1}) will lead to first-order partial differential equations in the dynamical variables $V^a$ and $A^{a}_{\ph{a}b}$ which do not look very familiar. In order to see how one may recover more familiar looking second-order differential equations we shall in this section introduce a `covariant' decomposition of the connection $A^{a}_{\ph{a}b}$. We first recall a simplifying strategy from the Einstein-Cartan theory of gravity where gravity is described
by the pair $\{\omega^{I}_{\ph{I}J},e^{I}\}$. However, in doing variations of actions it is convenient to make the following decomposition:
\begin{eqnarray}
\omega^{IJ}= \bar{\omega}^{IJ}+{\cal C}^{IJ} \label{vij}
\end{eqnarray}
where $\bar{\omega}^{IJ}$ is determined \emph{entirely} by $e^{I}$ and its partial derivatives via the following equation:

\begin{eqnarray}
D^{(\bar{\omega})}e^{I}=0. \label{toro}
\end{eqnarray}
which can be solved yielding
\begin{eqnarray}
\bar{\omega}^{IJ}_{\ph{ab}\mu} &=& 2(e^{-1})^{\nu[I}\partial_{[\mu}e^{J]}_{\nu]}+e_{\mu K}(e^{-1})^{\nu I}(e^{-1})^{\alpha J}
\partial_{[\alpha}e^{K}_{\nu]} \label{k1}
\end{eqnarray}
where $(e^{-1})^{\mu}_{I}$ is the inverse of the matrix $e^{I}_{\mu}$ referred to as the 
tetrad. By inspection $\bar{\omega}^{IJ}$ transforms as a gauge field and thus alone becomes responsible for the inhomogeneous transformation law of $\omega^{IJ}$; the one-form ${\cal C}^{IJ}$, referred to as the contorsion, transforms homogeneously under a gauge transformation \cite{Leclerc:2005ig,Hehl:2007bn}. 
The two-form $D^{(\omega)}e^{I}$ is referred to as the torsion and hence, as a solution to (\ref{toro}), $\bar{\omega}^{IJ}$ is often referred to as the torsion-free spin-connection. The equations of motion for the Einstein-Cartan theory  may be obtained by varying $e^{I}$ and ${\cal C}^{IJ}$ independently. A reason for why this is useful is that the equations of motion for ${\cal C}^{IJ}$ take a particularly simple form in Einstein-Cartan theory in the presence of the matter fields of the standard model of particle physics. Indeed, of these fields it is only spinorial matter that is expected to couple to the contorsion \cite{Hehl:1971qi}; the coupling is such that one can solve algebraically for the contorsion in terms
of spinorial current. This allows the contorsion to be reinserted into the action and eliminated from the variational principle. The resultant variational principle, dependent upon $e^{I}$ and $\bar{\omega}^{IJ}(e^{K})$, corresponds to the Einstein-Hilbert action and after the addition of the Gibbons-Hawking boundary term yields the Einstein equations for the metric tensor $g_{\mu\nu}$ following variation with respect to $e^{I}$.

We would like to make a similar decomposition in the Cartan gravity case. The idea again will be to encode the inhomogeneous transformation property of the connection in a quantity defined analogously to (\ref{toro}). The additional, homogeneously transforming terms in the connection will hopefully simplify analysis if some of them may be eliminated from the variational principle. 

Now consider the following decomposition of $A^{ab}$:

\begin{eqnarray}
A^{ab} &=& \bar{A}^{ab} + B^{ab} \label{split1}
\end{eqnarray}
As for the case of $\bar{\omega}^{IJ}$ in the Einstein-Cartan model, the field $\bar{A}^{ab}$ will be defined so as to encode the inhomogeneous transformation properties of the connection $A^{ab}$. %

The remaining part of the connection is the field $B^{ab}$, and by construction it transforms homogeneously with respect to $SO(1,4)|SO(2,3)$ gauge transformations. As follows, to simplify analysis we will assume that $V^{a}\neq 0$. From a dynamical perspective this restriction is not natural but we will see that the recovery of a four-dimensional metric in the theory is dependent the norm of $V^{a}$ being non-zero. Given this assumption, we may define the quantity $E^{a} \equiv B^{ab}V_{b}$ (this coincides precisely with the previously defined $E^{a}\equiv {\cal P}^{a}_{\ph{a}b}DV^{b}$). Furthermore, we may define the field $C^{ab}\equiv B^{ab}- (2/V^{2})E^{[a}V^{b]}$. We shall see that after symmetry breaking, $E^{a}$ plays the role of the co-tetrad $e^{I}$ whilst $C^{ab}$ plays the role of the contorsion ${\cal C}^{IJ}$. In this sense, the form $B^{ab}$ unifies co-tetrad and contorsion into a single object. For the null case where $V^{a}\neq 0$, $V^{2}=0$ we shall see that though the field $E^{a}$ is still well-defined, it corresponds to a 
metric tensor $\eta_{ab}E^{a}_{\mu}E^{b}_{\nu}$ which is three dimensional and therefore care is required.

We make the following ansatz for $\bar{A}^{ab}$, which is the contribution to $A^{ab}$ which transforms inhomogeneously under a local $SO(1,4)|SO(2,3)$ transformation:

\begin{eqnarray}
\label{wab}
\bar{A}^{ab}= -\frac{2}{V^{2}}dV^{[a}V^{b]}+\bar{\omega}^{ab}
\end{eqnarray}
where 

\begin{eqnarray}
\bar{\omega}^{ab}_{\ph{ab}\mu} &=& 2(E^{-1})^{\nu[a}{\cal P}^{b]}_{\ph{b]}c}\partial_{[\mu}E^{c}_{\nu]}+E_{\mu c}(E^{-1})^{\nu a}(E^{-1})^{\alpha b}
\partial_{[\alpha}E^{c}_{\nu]}
\end{eqnarray}
Where we have used the field $(E^{-1})^{\mu a}$, the `inverse' of $E^{a}$, which satisfies:

\begin{eqnarray}
V_{a}(E^{-1})^{a\mu}=0,\quad E^{a}_{\mu}(E^{-1})^{\mu}_{b}= {\mathcal P}^{a}_{\ph{a}c}{\mathcal P}^{c}_{\ph{c}b},\quad E_{a \mu}(E^{-1})^{a\nu}=\delta^{\nu}_{\mu}
\end{eqnarray}
By inspection the solutions for $\bar{\omega}^{IJ}$ and $\bar{\omega}^{ab}$ are identical up to interchange of $\{e^{I}_{\mu},(e^{-1})^{\mu}_{I}\}$ with $\{E^{a}_{\mu},(E^{-1})^{\mu}_{I}\}$. The notational coincidence is deliberate. As noted, after appropriate symmetry breaking induced by $V^{a}$ the field $E^{a}$ will play the roll of the co-tetrad $e^{I}$ and $\bar{\omega}^{ab}$ will play the roll of the torsion-free spin-connection $\bar{\omega}^{IJ}$. We will assume that $\bar{\omega}^{ab}$ is well defined and so the applicability of this variable will be restricted by the conditions this places upon $E^{a}$. The additional $V^{a}$-dependent 
contribution to $\bar{A}^{ab}$ encodes the inhomogeneously transforming part of $A^{ab}V_{b}$. By way of interpretation one may think of the five `degrees of freedom' of $V^{a}$ as being comprised of a norm $V^{2}$ and an orientation $U^{a}$ unit-vector (i.e. $U^{a}\equiv V^{a}/\sqrt{V^{2}}$, $|\eta_{ab}U^{a}U^{b}| = 1$). If $V^{2}> 0$ over a region of the manifold then one can choose
a gauge where $U^{a}\doteq \delta^{a}_{\ph{a}4}$. This fixes $\bar{A}_{ab}-\bar{\omega}_{ab}=0$ implying that $A^{ab}V_{b}\doteq E^{a}$. The following results (using equations (\ref{projector}), (\ref{wab}), and that $\bar\omega^{ab}V_b=0$) are helpful and may be considered `analogues' of the Einstein-Cartan vanishing-torsion equation $D^{\bar{\omega}}e^{I}=0$:
\begin{eqnarray}
D^{(\bar{A})}V^{a} &=&  dV^{a}+\bar{A}^{a}_{\ph{a}b}V^{b} =dV^{a}-\frac{1}{V^{2}}\left(dV^{a}V_{b}-dV_{b}V^{a}\right)V^{b}\nn\\
&=& \frac{1}{V^{2}}\left(V^{b}dV_{b}\right) V^{a}=\frac{1}{2V^{2}}dV^{2}V^{a} \label{c1}
\end{eqnarray}

\begin{eqnarray}
D^{(\bar{A})}E^{a} &=& dE^{a}-\frac{1}{V^{2}}\left(dV^{a}V_{b}-dV_{b}V^{a}\right)E^{b}+\bar{\omega}^{a}_{\ph{a}b}E^{b}=  dE^{a} -\frac{1}{V^{2}}V^{a}V_{b}dE^{b}+\bar{\omega}^{a}_{\ph{a}b}E^{b}\nn\\
&=&  dE^{a} -\frac{1}{V^{2}}V^{a}V_{b}dE^{b}-{\cal P}^{a}_{\ph{a}b}dE^{b}= dE^{a}-\eta^{a}_{\ph{a}b}dE^{b} = 0 \label{c2}
\end{eqnarray}
In summary then, we adopt the following decomposition of $A^{ab}$:

\begin{eqnarray}
A^{ab} &=& {\bar A}^{ab}(E^{c},V^{d})+ C^{ab} + \frac{2}{V^{2}}E^{[a}V^{b]} \label{aab}
\end{eqnarray}
Consequently if $V^{2}\neq0$ and $\bar{\omega}^{a}_{\ph{a}b}$ is well defined we may consider the action (\ref{act1}) as a functional of  of $\{V^{a},C^{ab},E^{a}\}$ complemented by the constraints on the projections of $C_{ab}V^{a}=E_{a}V^{a}=0$. Given the decomposition (\ref{aab}), we have

\begin{eqnarray}
F^{ab} &=& \bar{R}^{ab}-\frac{1}{V^{2}}E^{a}E^{b}+C^{a}_{\ph{a}c}C^{cb}+D^{(\bar{A})}C^{ab} + \frac{2}{V^{2}}V^{[b}C^{a]}_{\ph{a]}c}E^{c}+\frac{1}{V^{4}}E^{[a}V^{b]}dV^{2}\\
DV^{a} &=& \frac{1}{2V^{2}}dV^{2}V^{a}+E^{a}
\end{eqnarray}
where $\bar{R}^{ab}=d\bar{A}^{ab}+\bar{A}^{a}_{\ph{a}c}\bar{A}^{cb}$. We can now look to write the individual contributions to the action
$S[V^{a},A^{ab}]$ in terms of the variables $\{V^{a},C^{ab},E^{a}\}$. It will prove useful to group the contributions in terms of the manner in which $C^{ab}$ appears in the action. For some contributions terms with derivatives of $C^{ab}$ can be eliminated by partial integration resulting in an action with $C^{ab}$ entering only algebraically, whereas other terms are quadratic in derivatives of $C^{ab}$ so that the derivatives $D^{(\bar{A})}C^{ab}$ cannot be removed by partial integration. The former contributions (after partial integration) are as follows:
\begin{eqnarray}
S_{\alpha_{2}}[V^{a},E^{a},C^{ab}] &=& \int \left(\alpha_{2}E^{a}E^{b}C_{am}C^{m}_{\ph{m}b}+\frac{\alpha_{2}}{V^{2}}dV^{2}C_{ab}E^{a}E^{b}\right) \label{bein1}\\
S_{\beta_{1}}[V^{a},E^{a},C^{ab}] &=& \int \beta_{1}\epsilon_{abcde}V^{e}E^{a}E^{b}\left(\bar{R}^{cd}-\frac{1}{V^{2}}E^{c}E^{d}+C^{c}_{\ph{c}m}C^{md}\right) \nn\\
&&-\left(\frac{\partial\beta_{1}}{\partial V^{2}}+\frac{\beta_{1}}{2V^{2}}\right)\epsilon_{abcde}V^{e}dV^{2}C^{ab}E^{c}E^{d}\\
S_{\beta_{2}}[V^{a},E^{a},C^{ab}] &=& \int \frac{\beta_{2}}{2} dV^{2}C_{ab}E^{a}E^{b}\\
S_{\beta_{3}}[V^{a},E^{a},C^{ab}] &=& \int\left(\beta_{3}E^{a}E^{b}C_{am}C^{m}_{\ph{m}b}+\frac{\beta_{3}}{V^{2}}dV^{2}C_{ab}E^{a}E^{b}\right)\nn\\
&& -\frac{\partial\beta_{3}}{\partial V^{2}}dV^{2}C_{ab}E^{a}E^{b} \\
S_{\gamma_{1}}[V^{a},E^{a}] &=& \int \gamma_{1}\epsilon_{abcde}V^{e}E^{a}E^{b}E^{c}E^{d}
\end{eqnarray}
The remaining contributions, which involve derivatives of $C^{ab}$ are:
\begin{eqnarray}
S_{\alpha_{1}}[V^{a},E^{a},C^{ab}] &=& -2\int \frac{\alpha_{1}}{V^{2}}\epsilon_{abcde}V^{e}\left(\bar{R}^{ab}+C^{a}_{\ph{a}m}C^{mb}\right)E^{c}E^{d} +\int \alpha_{1}\epsilon_{abcde}V^{e}\left(\bar{R}^{ab}\bar{R}^{cd}+\frac{1}{V^{4}}E^{a}E^{b}E^{c}E^{d}\right)\nn \\
&& -2\left(\frac{\partial\alpha_{1}}{\partial V^{2}}+\frac{\alpha_{1}}{2V^{2}}\right)\epsilon_{abcde}V^{e}dV^{2}C^{ab}P^{cd}\\
S_{\alpha_{3}}[V^{a},E^{a},C^{ab}]&=& \int \alpha_{3}\bar{R}^{ab}\bar{R}_{ab} -2\frac{\partial\alpha_{3}}{\partial V^{2}}dV^{2}C_{cd}P^{cd} \label{bein2}
\end{eqnarray}
where
\begin{eqnarray}
P^{cd} \equiv \bar{R}^{cd}-\frac{1}{V^{2}}E^{c}E^{d}+\frac{1}{3}C^{c}_{\ph{c}m}C^{md}+\frac{1}{2}D^{(\bar{A})}C^{cd}
\end{eqnarray}
A detailed discussion of the derivation of equations (\ref{bein1}-\ref{bein2}) is given in \ref{odeto}. 

It is perhaps not immediately obvious why the equations resulting by variation with respect to $A^{ab}$ and $V^a$ would be equivalent from those obtained by varying with respect to $V^2$, $E^a$ and $C^{ab}$. One may worry that the ansatz/decomposition $A^{ab}$ (\ref{aab}) has led to terms with derivatives ${\cal P}^{a}_{\ph{a}c}D^{(A)}V^{c}={\cal P}^{a}_{\ph{a}c}\left(dV^{c}+A^{c}_{\ph{c}d}V^{d}\right)$ being represented by fields $E^{a}$ which appear algebraically. For example, if one would vary with respect to $F_{\mu\nu}$ for the action of a $U(1)$ gauge field one gets the incorrect equation $F_{\mu\nu}=0$. While the gauge field $U_\mu$ enters in the action with a derivative and so requires integration by parts when varying with respect to it, $F_{\mu\nu}$ enters algebraically with no integration by part necessary. Such a problem does not arise in our case for the reason that for some region where $V^{2}\neq 0$, the quantity ${\mathcal P}^{a}_{\ph{a}c}DV^{c}$ contains no gauge-invariant information about derivatives of $V^a$ as can be immediately verified in the specific gauge $V^a\doteq\phi\delta^a_4$, i.e. $E^a\doteq A^{ab}V_b$. Indeed, this is analogous to the case of a $U(1)$ Higgs model wherein one can write the complex Higgs field $\Phi$ in an arbitrary gauge as $\Phi=\chi(x^{\mu}) e^{ie\xi(x^{\mu})}$. One is also free to define the $U(1)$ gauge field $U_{\mu}= G_{\mu}- \partial_{\mu}\xi$, where $G_{\mu}$ is manifestly gauge-invariant. Writing the $U(1)$ Higgs action in terms of fields $\chi,\xi,G_{\mu}$, one finds that the phase field $\xi$ disappears from the action. Much like derivatives of part of the $U(1)$ Higgs field ($\xi$) have been used in the definition of the gauge field $U_{\mu}$, we have absorbed derivatives ${\cal P}^{a}_{\ph{a}c}dV^{c}$ into the connection $A^{ab}$ and as a result these derivatives, like those of $\xi$, do not appear in the action.

The distinction between the appearance of $C^{ab}$ in the actions is important. For any combination of the actions for which $C^{ab}$ appears algebraically (after a partial integration), the equation of motion for $C^{ab}$ may be used to solve for $C^{ab}$ itself. Consequently, this solution can be re-inserted back into the actions, yielding a variational principle that depends solely on $V^{a}$ and $E^{a}$. In this case the contorsion field does not have independent degrees of freedom and is instead completely fixed by other fields. On the other hand, whenever the action contains contributions from the $\alpha_1$ and $\alpha_3$ terms the contorsion may not be solved for algebraically and thus  must be regarded as a genuine dynamical field with independent degrees of freedom; i.e. we have propagating torsion.

The above seven actions remain completely general up to the requirement that $V^{2}\neq 0$ and that $\bar{\omega}^{ab}$ is well-defined. We will now concentrate on a sub-case which illustrates the consistency of the use of variables that assume $V^{2}\neq 0$. Though the approach of using variables which require a non-zero $V^{2}$ is clearly restrictive we will see that the variables enable in a simple way the recovery of more familiar looking geometric objects such as $g_{\mu\nu}$ and the torsion-free Ricci scalar ${\cal R}$. Indeed, this may be seen as a generalisation of the relationship between Einstein-Cartan gravity and the second-order formulation of gravity. The second-order formulation follows from the adoption of the ansatz (\ref{vij}) and does not admit more general solutions such as $e^{I}=0,\omega^{IJ}\doteq0$ that vacuum Einstein-Cartan theory does. Clearly though, the description of gravity using the Einstein-Hilbert action and its variable $g_{\mu\nu}$  is justified if solutions to the 
equation of motion are consistent with its assumed invertibility, even if it is regarded as only sometimes recoverable from the Palatini action of Einstein-Cartan theory.

We postpone discussion of the possibility that $V^{2}=0$ to Section \ref{vnull} but we stress that from the perspective wherein $V^a$ is a genuine dynamical field, {\em ad hoc} restrictions on $V^2$ (e.g. $V^2>0$, $V^2\neq0$, etc.) are rather unnatural. Instead, from the dynamical perspective we should in principle allow for possible solutions in which $V^2$ changes sign and can become zero on hyper surfaces. Such changes in the sign of $V^2$ will be accompanied with a change in metric signature. 
\section{Some useful notation}
Let us examine this a bit closer and also establish some notational conventions that will be used in the following sections. We will find that a metric formalism for gravity may be recovered and that  the signature of the metric tensor will depend on the sign of $V^{2}$ for a given group. For instance for $SO(1,4)$ if $V^{2}>0$ then the subgroup that leaves $V^{a}$ invariant is $SO(1,3)$ (corresponding to metric signature $(-,+,+,+)$). If $V^{2}<0$ then the subgroup is $SO(4)$ (metric signature $(+,+,+,+)$). For $SO(2,3)$ if $V^{2}>0$ then the subgroup is $SO(2,2)$ (metric signature $(-,-,+,+)$ and for $V^{2}<0$ 
the subgroup is $SO(1,3)$ (again corresponding to metric signature $(-,+,+,+)$). It is useful then to define a \emph{positive definite} scalar field $\phi\equiv |\sqrt{V^{2}}|$ such that

\begin{eqnarray}
\eta_{ab}V^{a}V^{b} =  \sigma \phi^{2}
\end{eqnarray}
The constant $\sigma=1$ for $SO(1,4)$ with $(-,+,+,+)$ metric signature and for $SO(2,3)$ with $(-,-,+,+)$ signature,
whilst $\sigma=-1$ for $SO(1,4)$ with $(+,+,+,+)$ signature and for $SO(2,3)$ with $(-,+,+,+)$ signature. Therefore in addition to keeping note of the norm of $V^{2}$ we must also keep track of the group being used as that will say what the metric signature is for a given $V^{2}$. Therefore we introduce the symbol $\theta$ which takes the value $-1$ if the metric has an even number of timelike dimensions (inclusive of the case that there are $0$) and the value $+1$ if the metric $g_{\mu\nu}$ has an odd number of timelike dimensions.

\section{A Specific Example}
\label{spece}
We are now ready to work out the physics of specific choices of actions and cast it in a more familiar second-order form in terms of the metric tensor $g_{\mu\nu}$. Consider then the case where, of the `$\alpha$,$\beta$,$\gamma$', only $\beta_{1}$,$\beta_{2}$ are non-vanishing. Furthermore we will assume that these two quantities are constant i.e. do not depend on $V^{2}$. As such $\beta_1$ and $\beta_2$ become constants with dimensions of $m^{-3}$ and $m^{-4}$ respectively. In the following $V^a$ and $E^a$ have dimensions $m$ and the remaining objects such as $A^{ab},C^{ab}, \bar{\cal R}^{ab},\dots$ dimensionless.

The combined action is then:
\begin{eqnarray}
S_{1}&=&\int \left(\beta_{1} \epsilon_{abcde}V^{e}+\beta_{2}V_{a}V_{c}\eta_{bd}\right)DV^{a}DV^{b}F^{cd}= \int \beta_{1}\epsilon_{abcde}V^{e}E^{a}E^{b}\left(\bar{R}^{cd}-\frac{1}{V^{2}}E^{c}E^{d}+C^{c}_{\ph{c}m}C^{md}\right)\nn\\
&&-\left(\frac{\beta_{1}}{2V^{2}}\epsilon_{abcde}V^{e}-\frac{\beta_{2}}{2}\eta_{ac}\eta_{bd}\right)dV^{2}C^{ab}E^{c}E^{d} \label{simact}
\end{eqnarray}
If $V^{a}$ is non-vanishing then it selects a subgroup which $V^{a}$ is invariant under. If we again utilize the orientation unit-vector $U^{a}\equiv V^{a}/|\sqrt{V^{2}}|$ then $\epsilon_{abcd} \equiv \epsilon_{abcde}U^{e}$ is invariant under the remnant transformations. It can be seen from (\ref{simact}) that variation of the action with respect to $C^{ab}$ will yield an equation linear in $C^{ab}$, which may be used to obtain the following solution for the field:
\begin{eqnarray}
C_{ab} &=& \frac{1}{2V^{2}}E_{[a}\partial_{b]}V^{2}+\frac{\beta_{2}\theta}{8\beta_{1}\phi}\epsilon_{abcd}\partial^{c}V^{2}E^{d}
\end{eqnarray}
where $\partial_{a}V^{2}\equiv (E^{-1})^{\mu}_{a}\partial_{\mu}V^{2}$. Substitution of $C^{ab}$ back into (\ref{simact}) yields \footnote{Substituting a solution back into the action is justified if the variable enters algebraically, possibly after integration by parts.}:
\begin{eqnarray}
\nn S_{1}[V^{2},E^{a}] &=&  \int \beta_{1}\epsilon_{abcd}\phi E^{a}E^{b}\bar{R}^{cd} -\beta_{1}\frac{1}{V^{2}}\epsilon_{abcd}\phi E^{a}E^{b}E^{c}E^{d}\\
&& +\epsilon_{abcd}\frac{\beta_{1}\phi}{16V^{4}}\left(1-\frac{\beta_{2}^{2}\theta\phi^{2}}{4\beta_{1}^{2}}\right)\left(6\partial_{m}V^{2}\partial^{b}V^{2}E^{a}E^{m}-\partial^{m}V^{2}\partial_{m}V^{2}E^{a}E^{b}\right)E^{c}E^{d} \label{s1dev}
\end{eqnarray}
We are now in a position to write $S_{1}$ in terms of more familiar variables. From $E^{a}$ we may define a metric tensor $\bar{g}_{\mu\nu} \equiv \eta_{ab}E^{a}_{\mu}E^{b}_{\nu}$ and inverse metric $\bar{g}^{\mu\nu}\equiv\eta_{ab}(E^{-1})^{a\mu}(E^{-1})^{b\nu}$. This compels us to identify $\bar{R}_{\mu\nu\alpha\beta}= E^{a}_{\mu}E^{b}_{\nu}{\bar R}_{ab\alpha\beta}$ with the torsionless Riemann curvature tensor. Furthermore we define the Ricci tensor $\bar{{\cal R}}^{\mu}_{\ph{\mu}\nu} \equiv \bar{R}^{\mu\alpha}_{\ph{\mu\alpha}\nu\alpha}$ and Ricci scalar $\bar{{\cal R}} \equiv \bar{{\cal R}}^{\mu}_{\ph{\mu}\mu}$. By standard methods, the action (\ref{s1dev}) may be rewritten as an integration over a scalar density:
\begin{eqnarray}\label{beforeredef}
S_{1}[\phi,\bar{g}_{\mu\nu}] &=& \int \beta_{1}\left(2\bar{{\cal R}}+\frac{3}{ \phi^{2}}\left(1-\frac{\beta_{2}^{2}\theta\phi^{2}}{4\beta_{1}^{2}}\right)\bar{g}^{\mu\nu}\partial_{\mu}\phi\partial_{\nu}\phi-\frac{24\sigma}{\phi^{2}}\right)\phi\sqrt{-\bar{g}}d^{4}x\nn\\
\end{eqnarray}

We may cast the action in a more familiar form by considering the conformal transformation of the metric: $g_{\mu\nu}=\Omega^2 \bar g_{\mu\nu}$. The Ricci scalar $\cal R$ corresponding to $g_{\mu\nu}$ is related to $\bar {\cal R}$ according to \cite{Wald:1984rg} $\bar{\cal R}= (\Omega^{2}{\cal R}+6g^{\mu\nu}\nabla_\mu\log\Omega\nabla_\nu\log\Omega+\bar{g}^{\mu\nu}\bar{\nabla}_\mu\bar{\nabla}_\nu\log\Omega)$. If we choose $\Omega^2=\phi/\phi_0$, for some constant $\phi_0$, the action in the new variables $(\phi,g_{\mu\nu})$ becomes

\begin{eqnarray}
\label{s1norm}
S_{1}[\phi,g_{\mu\nu}] &=& \int \left(2\beta_1\phi_0{\cal R}-\frac{3\phi_0\beta_{2}^{2}\theta}{4\beta_{1}}g^{\mu\nu}\partial_{\mu}\phi\partial_{\nu}\phi-24\beta_1\sigma\frac{\phi_0^2}{\phi^{3}}\right)\sqrt{-g}d^{4}x\nn\\
\end{eqnarray}
where the boundary term $\bar{\nabla}_\mu (6\beta_1\phi_0\bar{\nabla}^\mu\log\frac{\phi}{\phi_0})\sqrt{-\bar{g}}$ has been removed. We can now identify the gravitational constant as $\kappa=2\beta_1\phi_0=\frac{1}{16\pi G}$ where $G$ is Newton's constant. Furthermore, perhaps surprisingly, after this conformal transformation the `wrong-sign' kinetic contribution of equation (\ref{beforeredef}) is cancelled and the scalar field  $\phi^2=|V^2|$ is recognized as a scalar field satisfying a Klein-Gordon equation with potential $U(\phi)=24\beta_1\sigma\frac{\phi_0^2}{\phi^3}$.  Quite remarkably we see that the above simple action (\ref{simact}) corresponds to one of the more extensively studied models of dark energy in cosmology, namely the Peebles-Ratra rolling quintessence \cite{Ratra:1987rm,Peebles:1987ek}. These are scalar tensor theories with a potential of the form $\phi^{-\alpha}$. This highlights an important and novel connection between models of dark energy and Cartan gravity with dynamical symmetry breaking. 

Interestingly, as shown in \cite{Maroto:2006}, the value $\alpha=3$, which coincides with the model we have derived, remains consistent with supernovae data. However, the symmetry breaking field $V^a$ has non-trivial coupling to matter \cite{Westman:2012zk} which becomes important whenever $dV^2$ is significantly different from zero. Thus, the predictions of the Cartan-geometric Peeble-Ratra model are expected to be distinct from Peeble-Ratra models with no coupling to matter (such as those considered in \cite{Maroto:2006}).

The Peebles-Ratra potential does not have a minimum and does not allow for a stable  non-zero vacuum expectation value of $\phi$. In such a case there is no privileged choice of the value of $\phi_0$ which was introduced merely to make the conformal factor $\Omega$ dimensionless. However, if the potential could be modified by adding other polynomial terms to the action enabling a stable vacuum expectation value, then $\phi_0$ could be identified with that vacuum expectation value. How to achieve this will be considered below in Section \ref{stabil}.
\subsection{$SO(1,4)$ leading to $(-,+,+,+)$ metric signature}
Recall that for the group $SO(1,4)$ then if $V^{2} >0$ then the invariant subgroup is $SO(1,3)$ and with our conventions on $\eta_{ab}$ the metric $g_{\mu\nu}$ is of signature $(-,+,+,+)$. Therefore we have $\sigma=+1$ and $\theta=+1$. We can see from (\ref{s1norm}) that the action (\ref{simact}) corresponds to General Relativity coupled to a scalar field $\phi \equiv |\sqrt{V^{2}}|$ with right-sign kinetic term but with a potential with sign that depends upon the original gauge group. For $SO(1,4)$ we have a potential $1/\phi^{3}$. Therefore the field will tend to `roll down' the potential to increasing values of $\phi$. 

\subsection{$SO(2,3)$ leading to $(-,+,+,+)$ metric signature}

For the group $SO(2,3)$ if $V^{2}<0$ then the invariant subgroup is $SO(1,3)$ and with our conventions on $\eta_{ab}$ the metric $g_{\mu\nu}$ is of signature $(-,+,+,+)$, corresponding now to $\sigma=-1$ and $\theta=+1$. Now due to the different sign of $\sigma$ we have the potential $-1/\phi^{3}$. This potential will tend to lead to runaway evolution toward $\phi=0$. Therefore upon beginning from the action (\ref{simact}), we see that symmetry breaking is dynamically favoured for the case $SO(1,4)$.

\subsection{Stabilizing $\phi$}
\label{stabil}
We have seen that the action (\ref{simact}) is a model of dynamical symmetry breaking in the $SO(1,4)$ case. However, the potential for the field $\phi$ does not have a minimum, and the model in the absence of matter corresponds to a Peebles-Ratra rolling quintessence model \cite{Ratra:1987rm,Peebles:1987ek}. We now consider additional polynomial contributions to the action that would alter the potential as to create local maxima and minima. By inspection, aside from the $\alpha_{1}$ action it is only the $\gamma_{1}$ action that can provide additional contributions to the potential. All other terms involve the field $C^{ab}$ which generically will depend upon derivatives of $|V^{2}|=\phi^2$. It is straightforward to rewrite the $\gamma_{1}$ action in terms of the variables of the previous section:

\begin{eqnarray}
\label{s1stab}
S_{1}[\phi,g_{\mu\nu}] &=& \int \left(2\beta_1\phi_0{\cal R}-\frac{3\phi_0\beta_{2}^{2}\theta}{4\beta_{1}} g^{\mu\nu}\partial_{\mu}\phi\partial_{\nu}\phi -U(\phi)\right)\sqrt{-g}d^{4}x\nn\\
\end{eqnarray}
where we have defined the scalar field potential
\begin{eqnarray}\label{potential}
U(\phi) \equiv 24\phi_0^2\left(\frac{\beta_{1}\sigma}{\phi^{3}}-\frac{\gamma_{1}(\phi^{2})}{\phi}\right)
\end{eqnarray}
A local minimum in the potential corresponds to a real, positive solution to the equations $(dU/d\phi)_{\phi=\phi_{0}}=0$, $(d^{2}U/d\phi^{2})_{\phi=\phi_{0}} >0$. 
\begin{figure}
\centerline{\psfig{figure=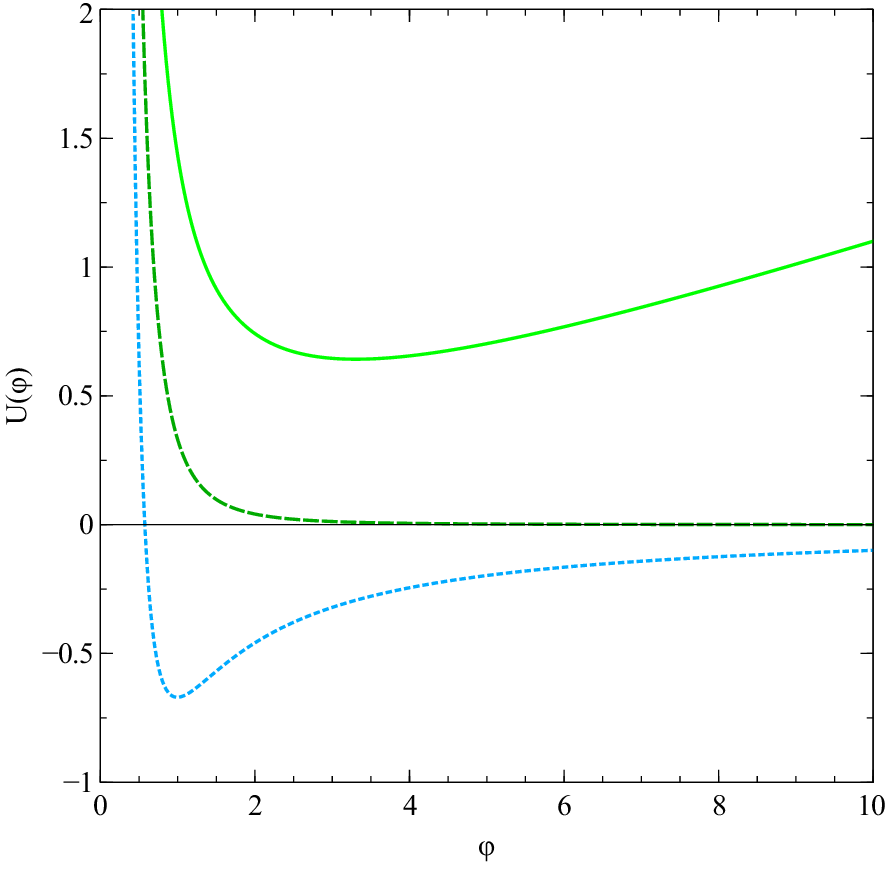,width=2.5truein}}
\caption{Some functional forms of $U(\varphi)$ (where $\varphi\propto\phi$, see equation (\ref{rescal})) for the $SO(1,4)$ case (units arbitrary). The dashed line represents the $+1/\phi^{3}$ `curvature'
contribution to the potential corresponding to Peebles-Ratra potential term. The dotted curve shows the recovery of a stable global minimum (negative cosmological constant if at this point) for an additional $\sim -1/\phi$ due to constant $\gamma_{1}$ contribution whilst the solid curve shows the recovery of a stable global minimum (positive cosmological constant if at this point) for an additional $\gamma_{1} \sim \mathrm{const.} + \phi^{2}$.}
\label{fig:so14}                       
\end{figure}

Note that a peculiarity of this approach is that appears impossible to add a term to $U(\phi)$ corresponding to a cosmological constant (i.e. a constant part of $U(\phi)$). This would, for instance, correspond to a contribution to $\gamma_{1}$ of the form $\sqrt{ \sigma V^{2}}$ which is excluded by the requirement that the action (\ref{act1}) is polynomial in $A^{ab}$ and $V^{a}$. However, if $\phi$ settles to a stationary point $\phi_{0}$ then it creates an effective cosmological constant. Furthermore the quadratic term in an expansion around $\phi_{0}$ defines a mass $M$ of the scalar field. 

\begin{figure}
\centerline{\psfig{figure=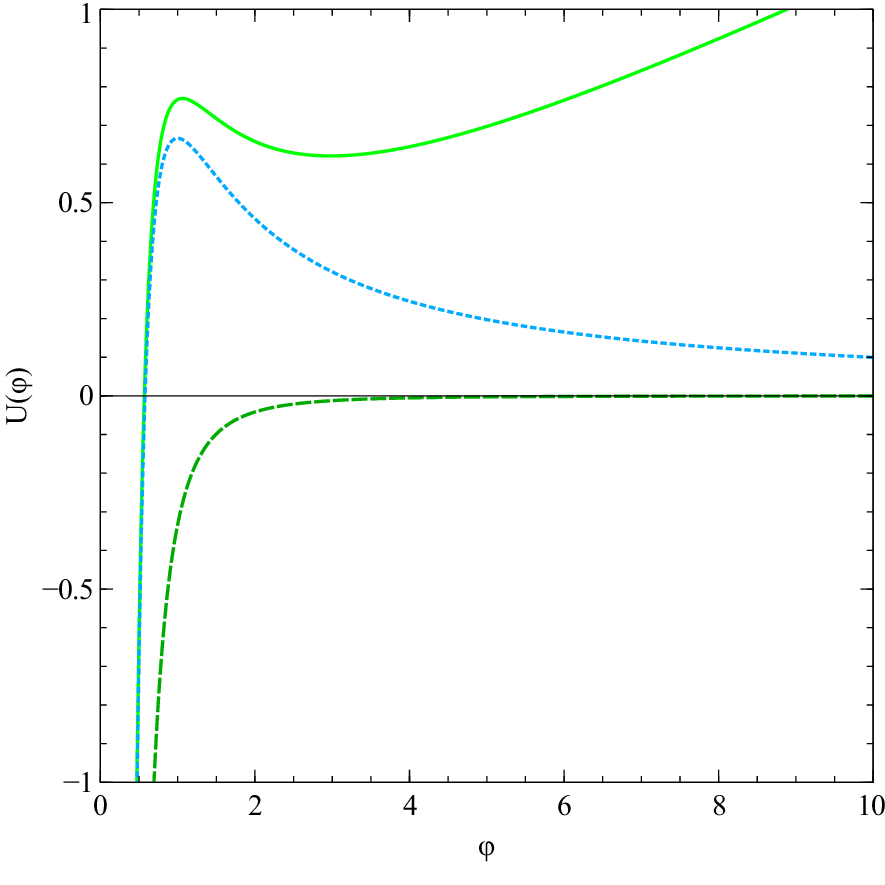,width=2.5truein}}
\caption{Some functional forms of $U(\varphi)$ for the $SO(2,3)$ case. The dashed line represents the $-1/\phi^{3}$ `curvature'
contribution to the potential. The dotted curve shows the recovery of an unstable maximum (positive cosmological constant if at this point) for an additional constant $\gamma_{1}$ contribution whilst the solid curve shows the recovery of a stable local minimum (positive cosmological constant) for an additional $\gamma_{1} \sim \mathrm{const.} + \phi^{2}$.}
\label{fig:so23}
\end{figure}
\subsection{An empirically viable model}
It is now interesting to examine whether there exists a specific choice of action which would lead to a modification of General Relativity yet be consistent with current data. In particular, such a model should presumably imply a positive cosmological constant at sufficiently low energies and avoid unwanted excitations of the scalar field. Of the six specific models displayed in Fig. 1 and 2 only one model seems free of potential or immediate problems. 

Clearly, the model corresponding to the dotted line of Fig. 1 is ruled out because of its stable negative value of the cosmological constant emulated by  $U(\phi_{min})$. We also note that all models based on $SO(2,3)$ seem potentially problematic as none of the potentials in Fig. 2 are bounded from below. This can potentially cause instability problems, especially in a quantum context with the possibility of quantum tunnelling. For this reason we will put them aside and regard them in absence of further analysis as unsuitable. Furthermore, the dashed line in Fig. 1 corresponds to an asymptotically zero cosmological constant ($U(\phi)\rightarrow 0$ as $\phi\rightarrow\infty$) and it is therefore not clear without further analysis that this model would be empirically viable. 

This leaves us with the model corresponding to the solid line in Fig. 1 which is characterized by having a global stable minimum of the potential $U(\phi)$. This model is based on the group $SO(1,4)$, a spacelike contact vector $V^a$, with $\beta_1$ and $\beta_2$ as constants, and with $\gamma_{1}$ possessing a single $V^{2}$-dependent correction, linear in $V^{2}$. We emphasise that a global minimum is guaranteed to exist when such a correction, if of the right sign, exists. More generally we may imagine $\gamma_{1}$ having a polynomial dependence on $V^{2}$ i.e.  $\gamma_{1} \equiv  -\Sigma_{n=0}^{\infty}\gamma_{1(2n)}V^{2n}$. The existence and distribution of minima in the potential $U(\phi)$ will depend on the set of constants $\gamma_{1(2n)}$. Though detailed considerations are beyond the scope of this work, it does not seem inconceivable that $\gamma_{1(2n)}\neq 0$ for some $n\geq 1$. The form of the scalar field kinetic term and potential for generic $\beta_{i}$ and $\gamma_{1}$ polynomial in $V^{2}
$ is considered in \ref{gencas}.

As follows we will assume that $\phi$ takes a value such that $\gamma_{1}$ is dominated by a quadratic correction i.e. $\gamma_{1} \sim -\gamma_{1(2)}V^{2}$ where $\gamma_{1(2)}$ is a constant. This assumption is for illustrative purposes, enabling us to simply relate the above constants to more familiar constants such as the gravitational and cosmological constants $\kappa$ and $\Lambda$ and an effective mass $M$ for the gravitational Higgs field $V^a$. 

The model is then defined by the action (\ref{s1stab}) with the scalar field potential
\begin{eqnarray}
U(\phi)=24\phi_0^2\left(\frac{\beta_1}{\phi^3}+\gamma_{1(2)}\phi\right)
\end{eqnarray}
corresponding to a choice of $\gamma(\phi^2)=-\gamma_{1(2)}\phi^2$ for some constant $\gamma_{1(2)}$ of dimensions $m^{-7}$. 

As mentioned above, the constant $\phi_0$ was introduced to make the conformal factor $\Omega=\frac{\phi}{\phi_0}$ dimensionless. As such it is arbitrary. However, it is natural to fix it by requiring it to coincide with the value of $\phi$ at the potential minimum, i.e. we impose
\begin{eqnarray}
0=\left.\frac{\partial U}{\partial \phi}\right|_{\phi=\phi_0}=24\phi_0^2\left(-\frac{3\beta_1}{\phi_0^4}+\gamma_{1(2)}\right)
\end{eqnarray}
which implies 
\begin{eqnarray}\label{VEV}
\phi_0^4=\frac{3\beta_1}{\gamma_{1(2)}}.
\end{eqnarray}
Furthermore, whenever the scalar field has settled to its vacuum expectation value $\phi_0$ it is constant everywhere and the corresponding value of the potential $U(\phi_0)$ will then emulate a cosmological constant $\Lambda$ defined by $\int \kappa({\cal R}-2\Lambda)\sqrt{-g}$. The gravitational constant must be identified as $\kappa=2\beta_1\phi_0$ and together with equation (\ref{VEV}) we obtain $\Lambda=\frac{U(\phi_0)}{2\kappa}=\frac{48\beta_1}{\kappa\phi_0}=\frac{24}{\phi_0^2}$. We can now express $\beta_1$ and $\gamma_{1(2)}$ in terms of $\Lambda$ and $\kappa$
\begin{eqnarray}
\beta_1=\frac{\kappa\sqrt{\Lambda}}{\sqrt{96}}\qquad \gamma_{1(2)}=\frac{\kappa\Lambda^{5/2}}{768\sqrt{6}}.
\end{eqnarray}
Thus, if the constants $\beta_1$ and $\gamma_{1(2)}$ are chosen properly we reproduce the predictions of General Relativity plus a cosmological constant whenever the scalar field is in its ground state. Again we note that in the absence of matter, a `genuine' cosmological constant truly independent of $\phi$ cannot be constructed- such a term could only arise from non-polynomial terms in the action (\ref{act1}) such as a contribution to $\gamma_{1}$ of the form $\sqrt{V^{2}}$.

At sufficiently high energies scalar field $V^2$ may get excited from its lowest energy state, the vacuum expectation value $\phi_0$. At which energy that happens is dictated by the effective mass $M$ of the scalar field. 

In order to determine the effective mass we need to put the scalar field kinetic term in a standard form. This is achieved by considering the following field rescaling:

\begin{eqnarray}\label{rescal}
\phi=a\varphi \qquad a^{-2}=\frac{3\phi_0\beta_2^2}{4\beta_1} 
\end{eqnarray} 
The action now takes on the standard form

\begin{eqnarray}
\bar S_{1}[\varphi,g_{\mu\nu}] &=& \int \left(\kappa{\cal R}- g^{\mu\nu}\partial_{\mu}\varphi\partial_{\nu}\varphi- U(\varphi)\right)\sqrt{-g}d^{4}x
\end{eqnarray}
%
with $U(\varphi)=24\phi_0^2(\frac{\beta_1}{a^3\varphi^3}+\gamma_{1(2)}a\varphi)$. 
The effective mass $M$ of the scalar field $\varphi$ is defined as ($\phi_0\equiv a\varphi_0$) :
\begin{eqnarray}
M^2\equiv\left.\frac{1}{2}\frac{\partial^2 U}{\partial \varphi^2}\right|_{\varphi=\varphi_0}=\frac{1}{2}24\phi_0^2 \frac{12\beta_1}{a^3\varphi_0^5}=\frac{\kappa^2\Lambda^3}{288\beta_2^2}
\end{eqnarray}
and we see that for given values of $\kappa$ and $\Lambda$ it is the parameter $\beta_2$ that determines the effective mass of the scalar field $\varphi$. 

We have now seen how General Relativity is recovered in the low energy limit from a gauge theory with dynamical symmetry breaking defined by the action 
\begin{eqnarray}
S[A^{a}_{\ph{a}b\mu},V^{a}] &=& \int \left(\beta_{1}\epsilon_{abcde}V^{e}+\beta_{2}V_{a}V_{c}\eta_{bd}\right)DV^{a}DV^{b}F^{cd}  - \gamma_{1}(V^{2})\epsilon_{abcde}V^{e}DV^{a}DV^{b}DV^{c}DV^{d}
\end{eqnarray}
containing the two basic ingredients of such a theory: a $SO(1,4)$  gauge field $A^{a}_{\ph{a}b\mu}$ and a Higgs field $V^a$ with norm $V^2$ subject to a Klein-Gordon equation. In particular, at the fundamental level neither metric nor co-tetrad are necessary for the formulation of this theory. We reiterate that the necessary condition for the existence of a stable minimum for the field $\phi$ is that $\gamma_{1}$ contain corrections polynomial in $V^{2}$ and of appropriate sign.  However, as will be the subject of a forthcoming paper \cite{Magueijo:2013yya}, a different and perhaps more elegant way of achieving symmetry breaking yielding a positive non-zero vacuum expectation value consists of replacing the $\beta_1$ term by the MacDowell-Mansouri $\alpha_1$ term. Thus, the action with only a $\alpha_1$ and $\beta_2$ terms implements a mechanism to dynamically recover a constant $V^{2}$ of appropriate sign.
\section{More General Cases}
\label{general}
It is natural to ask how general the results of the previous section are. In \ref{gencas} an identical calculation is performed for the case where all $\alpha,\beta,\gamma$ except $\alpha_{1}$ and $\alpha_{3}$ are non-zero and allowed to have a polynomial dependence upon $V^{2}$. As for the case considered in Section \ref{stabil}, if $V^{2}\neq 0$ then the action is equivalent to a metric theory of gravity coupled to a scalar field. By inspection one may again make a conformal transformation of the metric to recover the familiar Einstein-Hilbert Lagrangian. To recover a canonical kinetic term for the scalar field, it is necessary to perform a redefinition of the scalar field. 

Finally we consider the $\alpha_{1}$ and $\alpha_{3}$ terms. As mentioned, variation of these actions with respect to $C^{ab}$ will yield field equations containing derivatives of $C^{ab}$. Therefore the presence of these terms in an action generally precludes the possibility of eliminating $C^{ab}$ from the variational principle. In this sense these actions involve `propagating contorsion'. We now consider the impact of these terms.
\subsection{Effect of the $\alpha_{1}$ term}
We first consider the limiting case where $V^{2}$ is constant and non-vanishing. The action in this limit is commonly referred to as the Macdowell-Mansouri action \cite{MacDowell:1977jt}. Here it may be checked that the term $\epsilon_{abcde}V^{e}\bar{R}^{ab}\bar{R}^{cd}$ reduces to a boundary term whereas the remaining terms (i.e. those that do not vanish when $dV^{2}=0$) amount to Einstein-Cartan gravity in the presence of a cosmological constant. The Macdowell-Mansouri action has been studied in a wide variety of contexts \cite{Wise:2006sm,Wise:2011ab,Gibbons:2009af,Randono:2010cq,Gryb:2012qt,Obregon:2012zz,Durka:2012wd,Freidel:2012np}. When $V^{2}\neq 0$ and varying, a number of terms in $S_{\alpha_{1}}$ that are proportional to $dV^{2}$ may now contribute. This action has been considered briefly in \cite{Banados:1996gi} but this model, with the remnant $SO(1,3)$ symmetry of the symmetry broken phrase, seems largely unexplored. Cosmological solutions for a very closely related action to $S_{\alpha_{1}}$ 
have been recently considered \cite{TolozaT:2013wi}.

\subsection{Effect of the $\alpha_{3}$ term}
The structure of the $\alpha_{3}$ term is rather more familiar. Recall that in the event that $V^{2}$ is constant then $\alpha_{3}(V^{2})$ is necessarily constant and the term is simply a boundary term. If $V^{2}\neq 0$ and varying and $\alpha_{3}$ carries a dependence on $V^{2}$ then the $\alpha_{3}$ term corresponds to an additional term in gravitation in the Chern-Simons modified gravity theory \cite{Alexander:2008wi,Cantcheff:2008qn,Ertem:2009ur,Alexander:2009tp,Garfinkle:2010zx,Furtado:2010qv,Canizares:2012ji}, with $\alpha_{3}$ playing the role of the Chern-Simons scalar field $\Phi_{cs}$. Typically also present in these theories are an Einstein-Hilbert (or Palatini) action as well as an action for the dynamics of $\Phi_{cs}$. A direct comparison between Chern-Simons modified gravity a subset of the general actions (\ref{act1}) requires some care. We have seen that the $\beta_{1}$ action may reduce to the Einstein-Hilbert action only after conformal transformation of $E^{a}$, therefore a more accurate 
comparison follows after an additional conformal transformation of the $\alpha_{3}$ action. A persistent commonality though is the presence of derivatives of the contorsion, if allowed for, in the action of the theory. As in the case of the $\alpha_{1}$ term, the presence of a non-vanishing $\alpha_{3}(V^{2})$ may preclude the existence of a second-order formulation in terms of a metric and scalar field.

\section{Null or vanishing $V^{a}$}
\label{vnull}

It has become apparent in the previous sections that the tensor $\bar{g}_{\mu\nu}=\eta_{ac}V_{b}V_{d}B^{ab}_{\ph{ab}\mu}B^{cd}_{\ph{cd}\nu}= E^{a}_{\mu}E_{a \nu}$ is to be identified, up to a conformal factor, with the familiar metric tensor in situations where $V^{a}\neq 0$. However, dynamically it may be the case that $V^{a}$ becomes null (i.e. $\eta_{ab}V^{a}V^{b}=0$). This condition may be satisfied with $V^{a}=0$ or $V^{a}\neq 0$. When $V^{a}=0$ then $\bar{g}_{\mu\nu}=0$ and so it is unclear whether a metric description exists in regions where this is satisfied. The null case with $V^{a}\neq 0$ is a bit more subtle. In this case we may choose a gauge where $V^{a}=\psi(x^{\mu})(1,0,0,0,\pm 1)$, where first and last components refer to internal time and spatial components. The remaining components will be labelled by lowercase middle-alphabet Latin indices $i,j..$.

\begin{eqnarray}
\bar{g}_{\mu\nu}= \psi^{2}\eta_{ij}\left(B^{i}_{\ph{i}0} \pm B^{i}_{\ph{i}4}\right)_{\mu}\left(B^{j}_{\ph{j}0}\pm B^{j}_{\ph{j}4}\right)_{\nu}
\end{eqnarray}
Therefore in the null case the signature of the metric will be that of $\eta_{ij}$ i.e. the signature will be three dimensional rather than four dimensional. For $SO(1,4)$ then the metric has signature $(+,+,+)$ and for $SO(2,3)$ the metric will have signature $(-,+,+)$. The $\alpha_{1}$ action in isolation was studied for the former possibility in \cite{Westman:2012np}, wherein it was indeed found that a three dimensional geometric description was appropriate. Recall that for $SO(1,4)$ when $V^{2} >0$ then the metric is of signature $(-,+,+,+)$ whereas when $V^{2}<0$ the metric is of signature $(+,+,+,+)$. It is tempting to wonder whether situations may occur where the transition of $V^{2}$ through $0$ dynamically facilitates a change of metric signature. Such a possibility has been examined in the $(3+1)$ formulation of General Relativity \cite{Ellis:1991st}. Consideration of explicit models in Cartan Gravity, however, is beyond the scope of this current work. 
\section{Discussion and outlook}
\label{discussion}
A common view in physics has been that gravity should ultimately be thought of just another force field on par with the strong and electroweak fields of the standard model. From a mathematical perspective such a view is limited by the difference in mathematical machinery used by the force fields of particle physics on the one hand and the gravitational field on the other.  More specifically, while the force fields of particle physics are mathematically represented by gauge connections valued in some suitable Lie algebra, the gravitational field is typically described by a symmetric second rank metric tensor $g_{\mu\nu}$. The mathematical difference is still present within the Einstein-Cartan formulation of gravity in terms of a $SO(1,3)$  gauge connection $\omega^{IJ}$ and a co-tetrad $e^I$. In particular, the co-tetrad regarded as a fundamental object appears to have no analogue within ordinary gauge theory \footnote{An interesting alternative is in schemes that unify the $SO(1,3)$ gauge field with 
gauge fields of grand unified theories into a connection  ${\cal A}^{\Omega}_{\,\,\,\,\Gamma\mu}$ for a larger Lie group ${\cal G}$ \cite{Percacci:1984ai,Percacci:2009ij,Nesti:2009kk}. In these theories there is also taken to exist an object $\theta^{\Omega}_{\mu}$, which is a ${\cal G}$ vector and space-time one-form. It is this object which may display a symmetry breaking solution to play the role of the $SO(1,3)$  co-tetrad.}. 

However, it has been know for some time that gravity can be formulated exclusively in terms of objects of gauge theories with spontaneously broken symmetry. Within this Cartan-geometric formulation of gravity the descriptor of gravity consist of the pair $\{A^{a}_{\ph{a}b},V^{a}\}$. These objects admit a simple geometrical interpretation in terms of `idealized-waywisers' \cite{Westman:2012xk}, but from a particle physics point of view we recognize $A^{a}_{\ph{a}b \mu}$ as a standard gauge connection and $V^a$ as a symmetry breaking Higgs field. In this paper we have shown that there exist actions which are polynomial in these fields and their exterior derivatives which allow for a dynamical symmetry breaking mechanism closely mirroring that of the electroweak theory of particle physics. It is noteworthy that this is achieved without any metric tensor or co-tetrad appearing in the action. Instead, these objects are regarded as compound objects which can be constructed from the more fundamental variables $\{A^{
a}_{\ph{a}b},V^{a}\}$ should it be convenient to do so. In relation to this, we note that the constrained-norm form of Cartan gravity has been recovered from a perspective where $V^{A}$ itself is a composite object, regarded as a fermionic condensate constructed from either $spin(1,4)$ or $spin(2,3)$ spinor fields \cite{Randono:2010ym}.

To our knowledge the recovery of such symmetry breaking solutions following free variation of 
$\{A^{a}_{\ph{a}b},V^{a}\}$ in a polynomial action principle is new \footnote{There exist claims \cite{Chamseddine:1977ih} that a particular sub-class of polynomial actions (specifically a combination of $\alpha_{1}$, $\beta_{1}$, and $\gamma_{1}$ terms) yield a second-order scalar-tensor theory description for  small perturbations of freely varied fields $A^{ab}$ and $V^{a}$ around a Minkowski space geometry for $\bar{g}_{\mu\nu}$ in a symmetry broken-phase where $V^{a}=\phi(x^{\mu})\delta^{a}_{4}$. However, for these actions the resulting scalar excitation's kinetic term is removed entirely by conformal transformation to the frame in which the tensor perturbation is described by the Einstein-Hilbert action i.e. the perturbed system is that of General Relativity alongside a scalar field $\varphi$ appearing algebraically and coupling only to the determinant of the metric. }. By way of contrast, the constancy of $V^2$ is typically imposed by hand within Cartan gravity without any dynamical mechanism proposed. From a pure mathematical point of view we can of course regard $V^a$ as a representation of a preferred section on the gauge fibre bundle that is necessary for the mathematical construction of Cartan geometry \cite{Petti:2006ue}. On that view the norm $V^2$ does not play any role in the construction and we might as well impose $V^2=\mathrm{const.}$. However, from a particle physics point of view, and in particular with the electroweak Higgs field in mind, the restriction $V^2=\mathrm{const.}$ does not appear natural. Instead, it suggests a natural  modification of Cartan gravity in which the Higgs field $V^a$ is treated as a genuine dynamical field subject to equations of motion. This was the basic motivation for this paper. 

We found that one of the most extensively studied models of dark energy, the Peebles-Ratra rolling quintessence, is contained in disguise as the simple action (\ref{simact}). This establishes a novel connection between these specific models of dark energy and Cartan gravity with dynamical symmetry breaking. This paves the way for future research into whether dark energy can have a Cartan-geometric explanation.

Although the class of actions (\ref{act1}), which always lead to first-order partial differential equations, look rather unfamiliar they may (excluding $\alpha_1$- and $\alpha_3$-terms) be generically recast into a more familiar second-order form where the actions reduce to particular examples of scalar-tensor theories. The correspondence extends to the inclusion of $\alpha_{1}$ and $\alpha_{3}$ terms in a number of situations, for example for suitably small perturbations of $E^{a}$, $C^{ab}$, and $V^{a}$ around a locally Minkowski solution. Furthermore, we showed how a specific subset of (\ref{act1}) lead to a scalar potential with a global minimum playing the role of a vacuum expectation value. Agreement with General Relativity is guaranteed as long as the effective mass $M$ calculated from the potential is large enough to suppress the excitation of the Higgs field. Specifically, the existence of minima in the scalar potential requires the existence of corrections to a constant value of $\gamma_{1}$ that 
are polynomial in $V^{2}$. We are not aware of a reason why such terms should not generically be allowed for in the action (\ref{act1}) and so it is difficult to make substantive statements about whether the existence of a minima in the potential (with acceptable value of scalar field mass and cosmological constant) represents the need for `fine-tuning'. This remains an open question.

Interestingly the requirement that (\ref{act1}) is polynomial in basic variables precludes the possibility of a genuine `bare' cosmological constant term according to the volume form $\sqrt{-g}d^{4}x$. For high energies deviations from General Relativity are expected due to the presence of gravitational Higgs bosons as well coupling between Higgs field gradients and derivatives of the contorsion field in $\alpha_{1}$ and $\alpha_{3}$ terms.

One perhaps surprising result of this paper is that $\phi\equiv |\sqrt{V^{2}}|$ appears as a standard scalar field when the action is rewritten in a second-order form. From a differential forms point of view this shows that the Hodge dual pops up naturally. The natural appearance of the Hodge dual when rewriting polynomial first-order field equations in a second-order form was also noted in \cite{Westman:2012zk}. This seems to indicate that the use of auxiliary fields, see e.g. \cite{Smolin:2007rx,Lisi:2010td}, may not be necessary in order to reproduce equations with such a Hodge structure, e.g. the Klein-Gordon and Yang-Mills equations.

One central simplifying assumption in the analysis of this paper was that $V^2$ be everywhere non-zero. We have argued that this is akin to the use of an invertible metric $g_{\mu\nu}$ in the variational principle of second-order General Relativity, the use of which is consistent whenever solutions to the resultant equations of motion do not threaten the assumption of invertibility.
However, if we take the dynamical perspective seriously such a restriction on $V^{2}$ must be regarded as {\em ad hoc}. 
Instead, one should take the view that the field equations should dictate what possible solutions of $V^a$ we can have. As such we cannot exclude the possibility of $V^2$ changing sign or of $V^a$ even becoming zero on hypersurfaces. Under such conditions  the scalar-tensor formalism breaks down but perhaps not the applicability of the basic variables $\{A^{a}_{\ph{a}b},V^{a}\}$. As an interesting dynamical possibility we might have metric signature change. This would occur whenever $V^2$ changes sign, selecting out a different subgroup of $SO(1,4)|SO(2,3)$. Within a model based on $SO(1,4)$, $V^2$ may be spacelike in some region leading to the usual Lorentzian signature $(-,+,+,+)$ of the metric (based on subgroup $SO(1,3)$). However, it might be possible that the first-order field equations allow solutions where the sign of $V^2$ varies from region to region. In particular, in regions where $V^2$ is timelike, the `spacetime' becomes Euclidean with signature $(+,+,+,+)$ (based on subgroup $SO(4)$). It would 
therefore be interesting to investigate whether there are solutions to the first-order field  equations that admit a signature change and thus a dynamical and classical realization of Hartle and Hawking's no-boundary proposal \cite{Hartle:1983ai}.\footnote{ Indeed, the $\alpha_{1}$ term in isolation with FRW symmetry imposed straightforwardly yields cosmological solutions with signature change replacing the big bang singularity \cite{ZW2013}.} Indeed this represents a subtlety in this approach to viewing gravity as a gauge theory: changes
in the character of the symmetry breaking or even symmetry restoration inevitably coincide with the 
loss of familiar notions of space and time.

The present work may be taken to suggest a more radical view on the form of field equations in Nature. Typically second-order field equations are regarded as the ones chosen in Nature. In particular, gravity, Yang-Mills fields, and scalar fields are all described by second-order partial differential equations. A notable exception are fermionic fields described by Dirac equations which are naturally on first-order. Further, the Cartan-geometric formulation of gravity and the coupling to matter fields through the gauge prescription  \cite{Westman:2012zk} puts all fields equations on first-order form. Of course, this may be regarded as a mere reformulation. However, as we have noted in this paper, the inclusion of terms like $\alpha_1$ and $\alpha_3$ will not allow for an equivalent second-order scalar-tensor theory. With such contributions to the action the contorsion field becomes a genuinely dynamical field which cannot be solved for algebraically in terms of other fields. However, whenever the scalar $V^2$ becomes 
constant the contorsion field can be solved for algebraically and a second-order formulation becomes possible. From this perspective, we may perhaps regard the ubiquitousness of second-order field equations in nature as a contingent feature of a universe where the Higgs field $V^a$ has settled to its vacuum expectation value. However, at a more fundamental level the theory would be governed by first-order equations that cannot be cast into familiar second-order scalar-tensor theories.

As a further departure from the standard picture of gravity, it is conceivable that the $SO(1,4)|SO(2,3)$ invariance is itself part of a symmetry broken phase of a model with a larger symmetry. For instance, it may be interesting to construct an equivalent of the actions (\ref{act1}) for a gauge theory of the conformal group $SO(2,4)$; conceivably the dynamics of symmetry breaking may allow for a General Relativity limit, and more comprehensive differences at higher energies.

We thank Max Ba\~nados, Brian Dolan, Pedro Ferreira, Jo\~ao Magueijo, and Adolfo Toloza for insightful discussions. We also thank an anonymous referee, whose comments have led to a great deal of improvement of the manuscript. HW was supported by the Spanish MICINN/MINECO Project FIS2011-29287, the CAM research
consortium QUITEMAD S2009/ESP-1594, and the CSIC JAE-DOC 2011 program. TZ was supported by STFC grant ST/J000353/1

\appendix

\section{General Case}
\label{gencas}
It may be shown after rather lengthy calculation that the case where only $\alpha_{1}$ and $\alpha_{3}$ are assumed to be zero and all other variables are allowed to depend polynomially upon $V^{2}$ leads to the following second-order action upon elimination of $C^{ab}$:

\begin{eqnarray}
S_{2}[\phi,\bar{g}_{\mu\nu}] &=& \int \left(2\beta_{1}\bar{{\cal R}}-\bar{\xi}(\phi)\bar{g}^{\mu\nu}\partial_{\mu}\phi\partial_{\nu}\phi -\frac{24\left(\sigma\beta_{1}-\gamma_{1}\phi^{2}\right)}{\phi^{2}}\right)\phi\sqrt{-\bar{g}}d^{4}x
\end{eqnarray}
where

\begin{eqnarray}
\bar{\xi}(\phi) &=& 3\frac{\left(\theta Q^{2}-4\left(\frac{\partial(\beta_{1}\phi)}{\partial \phi^{2}}\right)^{2}+\frac{2\sigma (\alpha_{2}+\beta_{3})\theta}{\beta_{1}\phi}\left(\frac{\partial(\beta_{1}\phi)}{\partial \phi^{2}}\right)Q\right)}{\beta_{1}\left(1+\frac{\theta}{\phi^{2}}\left(\frac{\alpha_{2}+\beta_{3}}{2\beta_{1}}\right)^{2}\right)} \\
Q &\equiv & \frac{\sigma\alpha_{2}}{\phi^{2}}+\frac{\beta_{2}}{2}+\frac{\sigma\beta_{3}}{\phi^{2}}
\left(1-\frac{\phi^{2}}{\beta_{3}}\frac{\partial\beta_{3}}{\partial \phi^{2}}\right)
\end{eqnarray}
As before we may make a conformal transformation. If $\beta_{1}$ itself has a dependence upon $V^{2}$ we define we make the following conformal transformation $\bar{g}_{\mu\nu}=
(\kappa/2\beta_{1}\phi)g_{\mu\nu}$, in terms of which the action becomes:

\begin{eqnarray}
S_{2}[\phi,g_{\mu\nu}] &=& \int \left( \kappa{\cal R} -\xi(\phi)g^{\mu\nu}\partial_{\mu}\phi\partial_{\nu}\phi
-24\left(\frac{\kappa}{2\beta_{1}}\right)^{2}\left(\frac{\sigma\beta_{1}}{\phi^{3}}-\frac{\gamma_{1}}{\phi}\right)\right)\sqrt{-g}d^{4}x 
\end{eqnarray}
where

\begin{eqnarray}
\xi(\phi) \equiv \left(\frac{\kappa}{2\beta_{1}}\right)\left(\frac{12}{\beta_{1}}\left(\frac{\partial (\beta_{1}\phi)}{\partial \phi^{2}}\right)^{2}+\xi(\phi)\right)
\end{eqnarray}
where recall that the $\alpha_{i},\beta_{i},\gamma_{i}$ may depend polynomially on $\phi^{2}$ and the constant $\kappa \equiv 1/16\pi G$.

\section{Explicit calculation of actions}
\label{odeto}

The idea behind writing the actions (\ref{bein1})-(\ref{bein2}) in the form they appear in was to reduce, insofar as possible, the appearance of terms in $\db C^{cd}$. The appearance
of terms such as $C^{ab}\db C^{cd}$ that persist even after integration by parts and ignoring boundary terms implies that $C^{ab}$ can no longer generally be solved for algebraically in the equations of motion. It is important to isolate such terms as it is by solving for $C^{ab}$, when possible, that we may make contact with more familiar scalar-tensor theories. The following properties are of considerable use:

\begin{eqnarray}
D^{(\bar{A})}E^{a}=E^{a}V_{a}=C^{ab}V_{b}= \bar{R}^{ab}V_{b}=D^{(\bar{A})}C^{ab}V_{b}=0  \label{zeros}
\end{eqnarray}
We will discuss specific aspects of the form of the actions for $\alpha_{1}$,$\alpha_{2}$,$\alpha_{3}$ terms as they contain the technical details necessary for treatment of actions associated with the remaining terms. We first consider the $\alpha_{1}$ term:

\subsection{$\alpha_{1}$ Term}
The action for this term is given by:

\begin{eqnarray}
S_{\alpha_{1}} &=&\int \alpha_{1}\epsilon_{abcde}V^{e}F^{ab}F^{cd} \nn\\
&=&\int  \alpha_{1}\epsilon_{abcde}V^{e}\left(\bar{R}^{ab}-\frac{1}{V^{2}}E^{a}E^{b}+C^{a}_{\ph{a}f}C^{fb}+D^{(\bar{A})}C^{ab} + \frac{2}{V^{2}}V^{[b}C^{a]}_{\ph{a]}c}E^{c}+\frac{1}{V^{4}}E^{[a}V^{b]}dV^{2}\right)\nn \\
&&\cdot\left(\bar{R}^{cd}-\frac{1}{V^{2}}E^{c}E^{d}+C^{c}_{\ph{c}g}C^{gd}+D^{(\bar{A})}C^{cd} + \frac{2}{V^{2}}V^{[d}C^{c]}_{\ph{a]}f}E^{f}+\frac{1}{V^{4}}E^{[c}V^{d]}dV^{2}\right)\nn\\
&=&\int \alpha_{1} \epsilon_{abcde}V^{e}\left(\bar{R}^{ab}-\frac{1}{V^{2}}E^{a}E^{b}+C^{a}_{\ph{a}f}C^{fb}+ \frac{2}{V^{2}}V^{[b}C^{a]}_{\ph{a]}c}E^{c}+\frac{1}{V^{4}}E^{[a}V^{b]}dV^{2}\right)\nn \\
&&\cdot\left(\bar{R}^{cd}-\frac{1}{V^{2}}E^{c}E^{d}+C^{c}_{\ph{c}g}C^{gd}+ \frac{2}{V^{2}}V^{[d}C^{c]}_{\ph{a]}f}E^{f}+\frac{1}{V^{4}}E^{[c}V^{d]}dV^{2}\right)\nn \\
 && +2\alpha_{1}\epsilon_{abcde}V^{e}\left(\db C^{ab}F^{cd}-\frac{1}{2}\db C^{ab}\db C^{cd}\right)  \label{act7} 
 \end{eqnarray}
We have thus separated terms which involve $\db C^{ab}$ from those that are independent of it. 
As an aside we note that the action involves the following term, quartic in $C^{ab}$:

\begin{eqnarray}
\epsilon_{abcde}V^{e}C^{a}_{\ph{a}m}C^{mb}C^{c}_{\ph{c}n}C^{nd} \label{4c}
\end{eqnarray}
We will now show that this term is identically zero. For $V^{e}\neq 0$, we can enumerate such $C^{ab}$ indices by four numbers $\{0,1,2,3\}$ (corresponding to `directions' in the five dimensional $SO(1,4)|SO(2,3)$ internal vector space orthogonal to $V^{e}$) due to the fact that by definition $C^{ab}V_{b}=0$. Consequently, we see that every contribution to (\ref{4c}) will be a permutation of:

\begin{eqnarray}
C^{0}_{\ph{a}m}C^{m1}C^{2}_{\ph{c}n}C^{n3} &=& \left(C^{0}_{\ph{a}2}C^{21}+C^{0}_{\ph{a}3}C^{31}\right)\left(C^{2}_{\ph{c}0}C^{03}+C^{2}_{\ph{c}1}C^{13}\right)
\end{eqnarray}
We can see the due to the properties of the wedge-product, the above term is identically zero as it will inevitably involve two identical one-forms wedged together.

We now turn to terms in (\ref{act7}) involving $\db C^{ab}$. The general strategy will be to employ integration by parts to relate, up to a boundary term, terms involving this derivative
to terms that do not, for example:

\begin{eqnarray}
\int d\left(\alpha_{1}\epsilon_{abcde}V^{e}C^{ab}F^{cd}\right) &=& \int \db\left(\alpha_{1}\epsilon_{abcde}V^{e}\right)C^{ab}F^{cd} +\alpha_{1} \epsilon_{abcde}V^{e}\db C^{ab}F^{cd} \nn \\
&&  -\alpha_{1}\epsilon_{abcde}V^{e}C^{ab}\db F^{cd}
\end{eqnarray}
It is rather straightforward in general to evaluate the contributions from $\db F^{cd}$ but we will focus on two contributions that care. The first of these is the term $\alpha_{1}\epsilon_{abcde}V^{e}D^{(\bar{A})}C^{ab}C^{c}_{\ph{c}m}C^{md}$, which may be related to a boundary term as follows:

\begin{eqnarray}
\int d\left(\epsilon_{abcde}V^{e}\alpha_{1} C^{ab}C^{c}_{\phantom{a}p}C^{pd}\right) &=&  \int \epsilon_{abcde}\db\left(\alpha_{1}V^{e}\right)C^{ab}C_{ap}C^{p}_{\ph{p}b}+\epsilon_{abcde}\alpha_{1}V^{e}\left(\db C^{ab}\right)C^{c}_{\phantom{a}p}C^{pd}\nn\\
&&   - \epsilon_{abcde}\alpha_{1}V^{e}C^{ab}\left(\db C^{c}_{\phantom{a}p}\right)C^{pd}+\epsilon_{abcde}\alpha_{1}V^{e}C^{ab}C^{c}_{\phantom{a}p}\left(\db C^{pd}\right)\nn\\
&=&   \int \epsilon_{abcde}\db\left(\alpha_{1}V^{e}\right)C^{ab}C_{ap}C^{p}_{\ph{p}b}+\epsilon_{abcde}\alpha_{1}V^{e}\left(\db C^{ab}\right)C^{c}_{\phantom{a}p}C^{pd}\nn\\
&&   - 2\epsilon_{abcde}\alpha_{1}V^{e}C^{ab}\left(\db C^{c}_{\phantom{a}p}\right)C^{pd}\nn\\
&=&  \int \epsilon_{abcde}\db\left(\alpha_{1}V^{e}\right)C^{ab}C_{ap}C^{p}_{\ph{p}b}+3\epsilon_{abcde}\alpha_{1}V^{e}\left(\db C^{ab}\right)C^{c}_{\phantom{a}p}C^{pd}\nn\\
\end{eqnarray}
where we have used the identity

\begin{eqnarray}
\epsilon_{abcde}V^{e}\left(\db C^{ab}\right)C^{c}_{\phantom{a}p}C^{pd}=  -\epsilon_{abcde}V^{e}C^{ab}\left(\db C^{c}_{\phantom{a}p}\right)C^{pd} \label{id1}
\end{eqnarray}
To see why this identity holds, consider the contribution on the left hand side due to the arbitrarily chosen $\db C^{01}$:

\begin{eqnarray}
4\epsilon_{0123e}V^{e}\left(\db C^{01}\right)\left(C^{2}_{\phantom{a}0}C^{03}+C^{2}_{\phantom{a}1}C^{13}\right)
\end{eqnarray}
Contributions from the right-hand side are of the form:

\begin{eqnarray}
&& -\epsilon_{0abde}V^{e}C^{ab}\left(\db C^{01}\right)C_{1}^{\ph{p}d} -\epsilon_{ab1de}V^{e}C^{ab}\left(\db C^{10}\right)C_{0}^{\ph{p}d} \nn \\
&=&  -\epsilon_{0ab2e}V^{e}C^{ab}\left(\db C^{01}\right)C_{1}^{\ph{p}2} -\epsilon_{0ab3e}V^{e}C^{ab}\left(\db C^{01}\right)C_{1}^{\ph{p}3} \nn \\
&& -\epsilon_{ab12e}V^{e}C^{ab}\left(\db C^{10}\right)C_{0}^{\ph{p}2} -\epsilon_{ab13e}V^{e}C^{ab}\left(\db C^{10}\right)C_{0}^{\ph{p}3}\nn\\
&=&  -2\epsilon_{0132e}V^{e}C^{13}\left(\db C^{01}\right)C_{1}^{\ph{p}2} -2\epsilon_{0123e}V^{e}C^{12}\left(\db C^{01}\right)C_{1}^{\ph{p}3} \nn \\
&& -2\epsilon_{0312e}V^{e}C^{03}\left(\db C^{10}\right)C_{0}^{\ph{p}2} -2\epsilon_{0213e}V^{e}C^{02}\left(\db C^{10}\right)C_{0}^{\ph{p}3}\nn\\
&=& 4\epsilon_{0123e}V^{e}\left(\db C^{01}\right)\left(C^{2}_{\phantom{a}0}C^{03}+C^{2}_{\phantom{a}1}C^{13}\right)
  \end{eqnarray}
 An additional term that requires some care is the term $\epsilon_{abcde}V^{e}\alpha_{1}\db C^{ab}\db C^{cd}$. To express this in an alternative form consider the following boundary term:
 
 \begin{eqnarray}
\int d\left(\epsilon_{abcde}V^{e}\alpha_{1}C^{ab}\db C^{cd}\right) &=&  \int \epsilon_{abcde}\db\left(V^{e}\alpha_{1}\right)C^{ab}\db C^{cd} +\epsilon_{abcde}V^{e}\alpha_{1}\left(\db C^{ab}\right)\db C^{cd} \nn \\
   && +\epsilon_{abcde}V^{e}\alpha_{1}C^{ab}\left(\bar{R}^{c}_{\phantom{c}f}C^{fd}+\bar{R}^{d}_{\phantom{d}f}C^{cf}\right)
\end{eqnarray}
where in the final line we have used the expression relating $\db\db$ to $\bar{R}^{ab}$. Thus, up to the boundary term, we may express $\epsilon_{abcde}V^{e}\alpha_{1}\db C^{ab}\db C^{cd}$ in terms of three other terms. Importantly, the terms that are linear in $\bar{R}^{ab}$ and quadratic in $C^{ab}$ exactly cancel other terms of this nature appearing elsewhere in the action following the use of the identity

\begin{eqnarray}
\epsilon_{abcde}V^{e}\bar{R}^{ab}C^{c}_{\ph{c}f}C^{fd}=-\epsilon_{abcde}V^{e}\alpha_{1}C^{ab}\bar{R}^{c}_{\phantom{c}f}C^{fd}
\end{eqnarray}
which follows from the identity (\ref{id1}) upon substitution $\db C^{ab} \rightarrow \bar{R}^{ab}$.

\subsection{$\alpha_{2}$ Term}
We now evaluate the $\alpha_{2}$ term, described by the following action:

\begin{eqnarray}
S_{\alpha_{2}} &=&\int \alpha_{2}V_{a}V_{c}\eta_{bd}F^{ab}F^{cd}
\end{eqnarray}
Here we must evaluate the projection of the curvature $F^{ab}$ along $V_{a}$. Using the results of (\ref{zeros}) we have that:

\begin{eqnarray}
F^{ab}V_{a} = -\left(C^{b}_{\phantom{b}c}E^{c}+\frac{1}{2V^{2}}E^{b}dV^{2}\right)
\end{eqnarray}
The form of equation (\ref{bein1}) then follows.

\subsection{$\alpha_{3}$ Term}
Evaluation of the $\alpha_{3}$ term proceeds in the same manner to the $\alpha_{1}$ term following replacement $\alpha_{1}\epsilon_{abcde}V^{e}\rightarrow \alpha_{3}\eta_{ac}\eta_{bd}$. We note that, as in the $\alpha_{1}$ case, the $\alpha_{3}$ action involves the following term, quartic in $C^{ab}$:

\begin{eqnarray}
C_{a}^{\ph{a}n}C_{nb}C^{am}C_{m}^{\ph{m}b}
\end{eqnarray}
We may simply re-label indices $m\leftrightarrow n$ to yield:

\begin{eqnarray}
C_{a}^{\ph{a}m}C_{mb}C^{an}C_{n}^{\ph{n}b} &=& -C_{a}^{\ph{a}n}C_{mb}C^{am}C_{n}^{\ph{n}b}=-C_{a}^{\ph{a}n}C_{m}^{\ph{m}b}C^{am}C_{nb} -C_{a}^{\ph{a}n}C_{nb}C^{am}C_{m}^{\ph{m}b}
\end{eqnarray}
Therefore, as the quantity is equal to the minus of itself, it is identically zero.

We now consider terms in $\alpha_{3}\eta_{ac}\eta_{bd}\db C^{ab}F^{cd}$  which, as in the $\alpha_{1}$ case will be re-expressed by integration by parts. Again two terms require particular care.
The first of these is the term $\alpha_{3}\left(\db C^{ab}\right)C_{ap}C^{p}_{\ph{p}b}$, which may be related to a boundary term as follows:

\begin{eqnarray}
\int d\left(\alpha_{3}\eta_{ac}\eta_{bd} C^{ab}C^{c}_{\phantom{a}p}C^{pd}\right) &=&  \int d\alpha_{3}C^{ab}C_{ap}C^{p}_{\ph{p}b}+\alpha_{3}\left(\db C^{ab}\right)C_{ap}C^{p}_{\ph{p}b}\nn\\
&&   - \alpha_{3}C^{ap}\left(\db C_{ab}\right)C^{b}_{\ph{p}p}+\alpha_{3}C^{ab}C_{ap}\left(\db C^{p}_{\phantom{p}b}\right)\nn\\
&=& \int d\alpha_{3}C^{ab}C_{ap}C^{p}_{\ph{p}b}+3\alpha_{3}\left(\db C^{ab}\right)C_{ap}C^{p}_{\ph{p}b}
\end{eqnarray}
The second term that requires some care is the term $\alpha_{3}\db C_{ab}\db C^{ab}$. To express this in an alternative form consider the following boundary term:
 
 \begin{eqnarray}
\int d\left(\alpha_{3}C_{ab}\db C^{ab}\right) &=&  \int d\alpha_{3}C_{ab}\db C^{ab}+\alpha_{3}\left(\db C_{ab}\right)\db C^{ab} \nn \\
   && +\alpha_{3}C_{ab}\left(\bar{R}^{a}_{\phantom{a}c}C^{cb}+\bar{R}^{b}_{\phantom{b}c}C^{ac}\right)\nn
\end{eqnarray}
As in the $\alpha_{1}$ case, the terms that are linear in $\bar{R}^{ab}$ and quadratic in $C^{ab}$ exactly cancel other terms of this nature appearing elsewhere in the action.

\section{Bibliography}
\bibliographystyle{hunsrt}
\bibliography{references}

\end{document}